\documentclass[twocolumn,usenatbib]{mnras}
\usepackage{graphicx}
\usepackage{bookmark}
\usepackage{threeparttablex}
\usepackage{tabularx}
\usepackage[T1]{fontenc}
\usepackage{ae,aecompl}
\usepackage{pdflscape}
\usepackage{adjustbox}
\usepackage{subcaption}
\usepackage{rotating}
\usepackage{savesym}
\usepackage{amsmath}
\savesymbol{iint}
\usepackage{wasysym}
\usepackage{graphicx}	
\usepackage{amssymb}	
\usepackage{etoolbox}
\usepackage{threeparttable}
\usepackage{environ}
\usepackage{hyperref}
\usepackage{rotating}
\usepackage{pdflscape}
\usepackage{adjustbox}

\usepackage{soul}

\title{UOCS \thanks {UVIT Open Cluster Study}--VIII. UV Study of the open cluster NGC 2506 using \textit{ASTROSAT}}
\author[Panthi et al.]{
Anju Panthi,$^{1}$\thanks{p20190413@pilani.bits-pilani.ac.in} 
Kaushar Vaidya,$^{1}$ 
Vikrant Jadhav,$^{2}$ 
Khushboo K. Rao,$^{1}$
\newauthor
Annapurni Subramaniam,$^{3}$
Manan Agarwal$^{4}$
and Sindhu Pandey,$^{5}$
\\
$^{1}$ Department of Physics, Birla Institue of Technology and Science, Pilani, Rajasthan-333031, India.\\
$^{2}$ The Inter-University Centre for Astronomy and Astrophysics,Post Bag 4, Ganeshkhind, Pune, Maharashtra-411007, India.\\
$^{3}$ Indian Institute of Astrophysics, Sarjapur Road, Koramangala, Bangalore, India.\\
$^{4}$ Department of Physics and Kavli Institute for Astrophysics and Space Research, Massachusetts Institute of Technology, MA 02139, USA.\\
$^{5}$ Aryabhatta Research Institute of Observational Sciences, Manora Peak, Nainital, India.}

\date{Accepted}

\pubyear{2022}

\begin{document}
\label{firstpage}
\pagerange{\pageref{firstpage}--\pageref{lastpage}}
\maketitle

\begin{abstract}
We study an intermediate-age open cluster NGC 2506 using the \textit{ASTROSAT}/UVIT data and other archival data. We identified 2175 cluster members using a machine learning-based algorithm, ML--MOC, on Gaia EDR3 data. Among the cluster members detected in UVIT filters, F148W, F154W, and F169M, we detect 9 blue straggler stars (BSS), 3 yellow straggler stars (YSS) and 3 red clump (RC) stars. We construct multi-wavelength spectral energy distributions (SEDs) of these objects to characterize them and to estimate their parameters. We discovered hot companions to 3 BSS, 2 YSS and 3 RC candidates and estimated their properties. The hot companions with estimated temperatures, T$\mathrm{_{eff}}$ $\sim$ 13250--31000 K, are WDs of extremely low-mass ($\sim$ 0.20 M$_\odot$), low-mass ($\sim$ 0.20--0.40 M$_\odot$), normal mass ($\sim$ 0.40--0.60 M$_\odot$), and high-mass ($\sim$ 0.8 M$_\odot$). We suggest that systems with extremely low mass and low mass WDs as companions are formed via Case-A/Case-B mass transfer mechanism. A BSS is the likely progenitor of the high mass WD, as a star with more than the turn-off mass of the cluster is needed to form a high mass WD. Thus, systems with high mass WD are likely to be formed through merger in triple systems. We conclude that mass transfer as well as merger pathways of BSS formation are present in this cluster. 
\end{abstract}


\section{Introduction} \label{Introduction}

Star clusters are ideal laboratories to study stellar populations in the host galaxy. Being a homogenous collection of stars having the same age, distances, kinematics, and metallicities, they provide the means to study the single and binary evolution of stars.
Blue straggler stars (BSS) are unusual stellar populations that are brighter and bluer with respect to the main sequence turnoff (MSTO) in the color-magnitude diagrams (CMD) of star clusters \citep{sandage1953color}. They are found in different stellar environments such as open clusters (OCs, \citealt{ahumada1995catalogue}), globular clusters (GCs, \citealt{sandage1953color}), Galactic fields \citep{preston2000these}, and dwarf galaxies \citep{momany2007blue}. BSS in star clusters are linked to the presence of close binary or multiple stellar systems, which are often formed through internal binary evolution or during the dynamical interaction between binaries and other stars \citep{knigge2009binary,mathieu2009binary,leigh2011analytic}.
Observational evidences show that BSS are among the most massive members of star clusters \citep{shara1997first, gilliland1998oscillating, ferraro2006discovery, beccari2006population, fiorentino2014blue} and therefore they tend to get concentrated toward the cluster centre as the cluster evolves \citep{ferraro1992galactic,ferraro2012dynamical,vaidya2020blue,2021MNRAS.508.4919R}. Due to this, BSS are considered crucial probes to study the interplay between stellar evolution and stellar dynamics \citep{bailyn1995blue}.

The three processes, namely direct stellar collision \citep{chatterjee2013stellar}, mass transfer in a binary system \citep{mccrea1964extended}, and merging in hierarchical triple stars \citep{perets2009triple}, are anticipated to be the primary ways in which BSS are produced. Stellar collisions are likely to occur in the environments where stellar density is quite high, such as the cores of GCs \citep{chatterjee2013stellar,hypki2013mocca,hurley2005complete}. The mass transfer mechanism may be further classified into three categories depending upon the evolutionary stage of the primary star when the mass transfer happens: Case-A, in which the primary is in the main sequence (MS) \citep{webbink1976evolution}, Case-B, in which the primary is in the red-giant branch (RGB) phase  \citep{mccrea1964extended}, and Case-C, in which the primary is in the asymptotic giant branch (AGB) phase \citep {chen2008binary}. 
The Case-A mass-transfer channel leaves behind either a single BSS or a binary BSS with a short period main-sequence companion, Case-B mass-transfer channel produces a short period (< 100 days) binary BSS with He WD as a companion, whereas Case-C mass transfer channel results in a long period binary (> 1000 days) BSS with CO WD as a companion. In the case of merger in a hierarchical triple system, the dynamical evolution of the triples through the Kozai mechanism and tidal friction can induce the formation of very close inner binaries. Angular momentum loss in a magnetized wind or stellar evolution could then lead to the merger of these binaries (or to mass transfer between them) and produce BSS in long binary (or triple) systems \citep{kiseleva1998tidal,fabrycky2007shrinking}.

Identification and characterization of BSS in OCs and GCs to infer their formation mechanism continues to be an important topic of research. Ultraviolet (UV) wavelengths are particularly suitable to study BSS, as BSS are much brighter in UV wavelengths owing to their relatively higher temperatures. Thus, in the UV wavelengths, the BSS define a clean sequence that is easily distinguishable in the CMD \citep{siegel2014swift, ferraro2018hubble}. A study illustrating the above by \cite{sahu2018uvit} found that BSS sequence stands out in the UV CMDs as compared to the optical CMDs. Additionally, UV-based studies allow the possibility of identifying BSS binaries with a hot companion from the excess flux in the UV wavelengths. \cite{knigge2000far} identified a BSS-white dwarf (WD) binary in 47 Tuc obtained with the \textit{Hubble Space Telescope (HST)} observations. In the OC NGC 188, \cite{gosnell2015implications} discovered WD companions of 7 BSS using \textit{HST} observations in far-ultraviolet (FUV) filters. For the same cluster, \cite{subramaniam2016hot} discovered a post AGB/HB companion of a BSS (WOCS 5885) using \textit{ASTROSAT}/Ultraviolet Imaging Telescope (UVIT) data. Similarly, \citet{sindhu2019uvit}, \citet{pandey2021uocs}, \citet{jadhav2021uocs}, and \citet{vaidya2022uocs} used UVIT data to study the OCs M67, King 2, and NGC 7789, respectively, and reported BSS with hot companions. UV observations also detect other interesting objects in star clusters, such as the yellow stragglers stars (YSS) and red clump (RC) stars. YSS are found blue-ward of the RGB and above the sub-giant branch in the optical CMDs \citep{stryker1993blue,clark2004blue}.
These are significantly brighter than the main sequence, but redder than the blue stragglers \citep{landsman1997s1040,leiner2016k2}, and are potentially the blue stragglers that have evolved into the giant or sub-giant stars. Recently, YSS were found in an OC M67 using the \textit{ASTROSAT}/UVIT data \citep{sindhu2019uvit}. The RC stars are cool horizontal branch stars, which have undergone a helium flash and are now fusing helium in their cores. They appear red and close to the RGB \citep{2016ARA&A..54...95G}.

NGC 2506 ($\alpha$ = $8^{h}$ $00^{m}$ $1.0^{s}$ , $\delta  = -10^{\circ} 46\arcmin 12\arcsec $) 
is an intermediate age ($\sim$ 2 Gyr) open cluster. Several photometric and spectroscopic studies have been done to estimate cluster  parameters including age, distance, reddening, and metallicity which are listed in Table \ref{table1}. Moreover, the BSS of this cluster have been identified in several studies \citep{xin2005blue,arentoft2007oscillating,knudstrup2020extremely,vaidya2020blue,jadhav2021blue,rain2021new}, but they have not been studied so far using the UV-wavelengths. The formation mechanisms of these BSS have also not been investigated earlier. We present the first UV-led multi-wavelength study of the BSS populations in NGC 2506 using the \textit{ASTROSAT}/ UVIT data. In addition to the BSS, we characterize the YSS and the RC stars of the cluster using multiwavelength data.

\begin{table*}
\caption{Parameters of NGC 2506 from the literature.}
\begin{tabular}{cccc}
\hline
Age&Distance&E(B-V)&[Fe/H]\\ 
(Gyr)&(pc)&&\\
\hline
\\ 
1.1--3.4 & 3110--3880&0.04--0.08 & $-$ 0.52--$-$0.19
\\
\hline
\label{table1}
\end{tabular}
\begin{tablenotes}
\item {Age : \citet{mcclure1981old, xin2005blue, anthony2016wiyn, vaidya2020blue, knudstrup2020extremely}}
\item { Distance : \citet{kharchenko2013global, rangwal2019astrometric, vaidya2020blue, knudstrup2020extremely}}
\item { E(B-V): \citet{mcclure1981old, kim2001search, carretta2004iron, xin2005blue, anthony2016wiyn, knudstrup2020extremely}}
\item { [Fe/H] : \citet{friel1993metallicities, reddy2012comprehensive,anthony2018wiyn,knudstrup2020extremely}} 
\end{tablenotes}
\end{table*}

The paper is arranged in the following manner: \S \ref{Section 2} describes the observations and data reduction procedure, \S \ref{Section 3} gives the data analysis, \S \ref{Section 4} has results and discussions, and \S \ref{Section 5} gives the conclusions of the work.

\section{Observations and Data reduction} \label{Section 2}

UVIT is one of the payloads in \textit {ASTROSAT} and has two 38 cm telescopes. One of the telescopes carries a far-UV (FUV) channel (130 -- 180 nm), and the other carries both a near-UV (NUV) channel (200 -- 300 nm) and a visible (VIS) channel (350 -- 550 nm). UVIT can perform simultaneous observations in these three channels in a circular field of view of diameter $\sim$ 28 $\arcmin$. It has a spatial resolution (FWHM) of $\sim1.2 \arcsec $ for the NUV filter and $\sim 1.5\arcsec$ for the FUV filter. The peak effective area excluding the losses in the filters used for band-selection of UVIT is $\sim$ 10 cm$^{2}$ for the FUV and $\sim$ 50 cm$^{2}$ for the NUV filter. The details of the UVIT instrument and calibration can be found in \cite{kumar2012ultraviolet}, \cite{subramaniam2016orbit}, and \cite{tandon2017orbit}. NGC 2506 was observed on 7$^{th}$ October 2019, under the \textit {ASTROSAT} proposal A07-005.

\begin{figure*}
\centering
\includegraphics[width = 10cm]{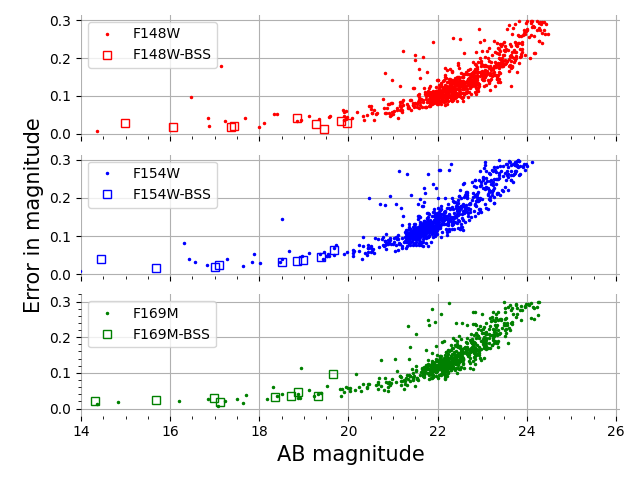}
\caption{PSF magnitudes vs. errors in magnitudes in different UVIT filters. Open squares denote the BSS in respective filters.}
\label{Fig. 1}
\end{figure*}

\begin{table*}
\caption{The details of \textit{ASTROSAT}/UVIT observations, source detections, and \textit{Gaia} EDR3 counterparts that are identified as members.} 
\adjustbox{max width=\textwidth}{
\begin{tabular}{ccccc}
\hline
\\
Filter Name&Exposure time (sec)&FWHM ($\arcsec$)&Detections& \textit{Gaia} EDR3 counterparts
\\
\hline
F148W (CaF2-1)&9224&1.07&776&464\\  
F154W (BaF2)&7499&1.09&740&438\\   
F169M (Sapphire)&7027&1.11&656&418 \\ 
\\
\hline
\label{table2}
\end{tabular}
}
\end{table*}

We obtained the science ready images from level 1 data available in the \textit {ASTROSAT} archive by doing distortion correction, flat field correction, and spacecraft drift using CCDLAB \citep{postma2017ccdlab,postma2021uvit}. The final science ready images in three FUV filters F148W, F154W, and F169M have exposure times of 9224, 7499, and 7027 seconds, respectively. These images were good for further analysis with an FWHM of $\sim$1 $\arcsec$ as mentioned in Table \ref{table2}. We performed the point-spread function (PSF) photometry on all the three UVIT images using the DAOPHOT package in IRAF \citep{stetson1987daophot}. The UVIT magnitudes were obtained in the AB magnitude system by using the zero-point (ZP) magnitudes given in \cite{tandon2020additional}. The aperture correction value in each filter was estimated using the curve of growth analysis technique and was applied to the PSF magnitudes. We also applied the saturation corrections to the magnitudes following \cite{tandon2020additional}. The magnitudes and errors of detected sources are shown in Figure \ref{Fig. 1}.

\section{Data analysis} \label{Section 3}

\subsection{Cluster membership}
In order to identify the members of NGC 2506, we used the machine-learning based algorithm, ML--MOC, developed by \cite{agarwal2021ml} on \textit {Gaia} EDR3 data  \citep{2021A&A...649A...1G}. The steps followed for determining the members of NGC 2506 using ML--MOC are briefly described here. All the sources with five astrometric parameters (RA, DEC, proper motions in RA, proper motions in DEC, and parallax) and with valid measurements in three Gaia photometric passbands G, G$_{\text{BP}}$, and G$_{\text{BP}}$, were classified as \textit{All sources} if their parallax values were non-negative and the errors in the G magnitudes were smaller than 0.005. From these sources, probable field stars were removed using the k-nearest neighbour (kNN, \citealt{cover1967nearest}) algorithm. These sources with a higher number of cluster members than field stars were termed as \textit{sample sources}. Then the Gaussian mixture model (GMM, \citealt{peel2000robust}) was applied to separate cluster and field members by fitting two Gaussian distributions in the proper motion and parallax space to the \textit{sample sources}. The membership probabilities of all the sources were also assigned using the GMM. We identified 2175 members with G < 20 mag within 30$\arcmin$ of the centre of NGC 2506. By cross-matching UVIT detected sources with these \textit {Gaia} EDR3 members within 1$\arcsec$ search radius using TOPCAT \citep{taylor2011topcat}, we found counterparts in all the filters. The information on exposure times of observations, the FWHM of sources, the number of detections in all the UVIT filters, and the \textit{Gaia} EDR3 members counterparts are given in Table \ref{table2}. To identify BSS of NGC 2506, we applied the methodology adopted by \cite{2021MNRAS.508.4919R} for segregating BSS from MSTO stars and binary stars located above the MSTO.
This methodology is briefly summarised as follows. We plotted a PARSEC isochrone \citep{bressan2012parsec} with suitable metallicity and age as given in Table \ref{table1} to the cluster CMD. We next
plotted the equal mass isochrone to isolate the binary stars. Finally, we selected the stars bluer than this isochrone as BSS. We thus identified 9 BSS in this cluster.

\begin{figure*}
\includegraphics[width=0.6\textwidth]{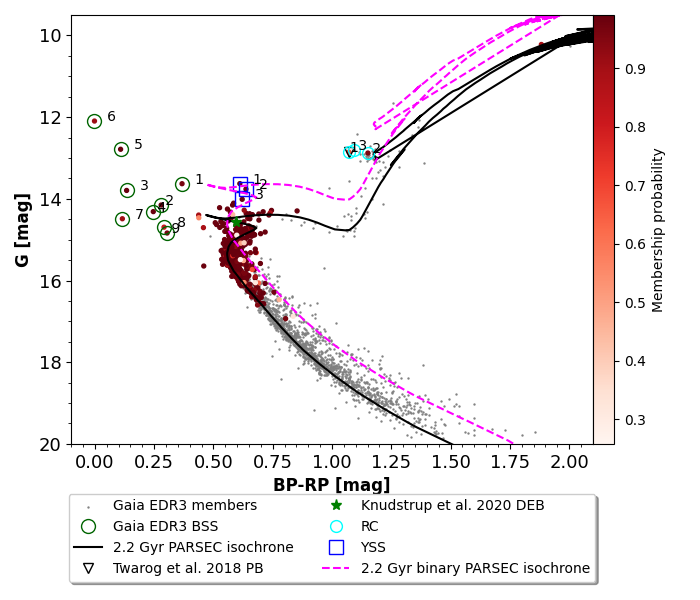} \par
\includegraphics[width=0.6\textwidth]{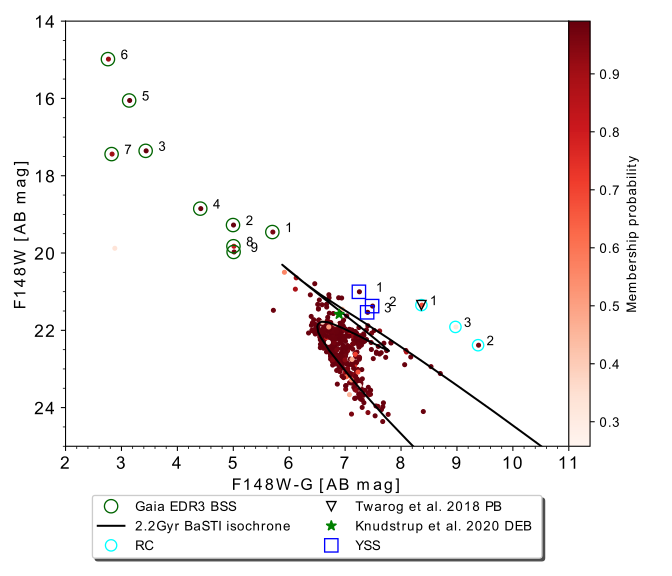}
\caption{The top figure shows the optical CMD of NGC 2506 showing all \textit {Gaia} EDR3 members as grey dots, BSS according to \textit {Gaia} EDR3 members as green open circles (labelled according to the coordinates given in Table \ref{table: Table 3}), F148W detections as red dots, YSS detected in all the UVIT filters as blue open squares, a detached eclipsing binary according to \citet{knudstrup2020extremely} as green star, a probable binary according to \citet{anthony2018wiyn} as black open triangle, and RC stars as cyan open circles. A PARSEC isochrone \citep{bressan2012parsec} of age = 2.2 Gyr, distance = 3110 pc, and Z = 0.004 is plotted after applying the extinction correction of A$_{\text{G}}$ = 0.39 and reddening of E(B$_{\text{P}}$ $-$ R$_{\text{P}}$) = 0.155. A binary sequence isochrone for the equal mass binaries with a G magnitude brighter than 0.75 mag than that of main-sequence is shown in pink dashed line. The bottom figure shows the UV-optical CMD showing all F148W detections in red dots. All other sources are represented with the same symbols as in the \textit {Gaia} EDR3 CMD. A BaSTI isochrone \citep{hidalgo2018updated} of the same cluster parameters as the \textit {Gaia} EDR3 CMD has been plotted after applying the extinction correction A$_{\text{F148W}}$ = 0.721, and reddening correction E(F148W $-$ G) = 0.473.}
\label{Fig. 2}
\end{figure*}

\subsection{The Colour-Magnitude Diagrams} \label{Section 3.2}
Figure \ref{Fig. 2} shows optical and UV CMDs of the cluster. In the optical CMD, Gaia EDR3 members including BSS and sources detected in the UVIT/F148W are shown. A PARSEC isochrone\footnote{http://stev.oapd.inaf.it/cgi-bin/cmd}\citep{bressan2012parsec} of age 2.2 Gyr, distance = 3110 pc, and Z = 0.0045 is over plotted after applying extinction correction A$_{\text{G}}$ = 0.39 and reddening correction E(B$_{\text{P}}$ $-$ R$_{\text{P}}$) = 0.155. The age and distance are taken from \cite{vaidya2020blue}. A binary sequence isochrone is also plotted in the optical CMD for the equal mass binaries with a G magnitude brighter than 0.75 mag than that of main-sequence stars, as shown by the pink dashed line.
The UV-optical CMD shows all the sources detected in UVIT/F148W filter that are also \textit{Gaia} EDR3 members. A BaSTI isochrone\footnote{http://basti-iac.oa-abruzzo.inaf.it/hbmodels.html} \citep{hidalgo2018updated} of the same fundamental parameters after applying extinction correction A$_{\text{F148W}}$ = 0.721 and reddening correction E(F148W$-$G) = 0.473 has been over plotted. The 3 YSS candidates detected in UVIT/F148W filter are represented by blue open squares, whereas 3 RC stars are marked as cyan open circles in both optical and UV CMDs. We have also shown the membership probabilities of identified members above G = 17 mag according to ML--MOC in both optical and UV CMD. All the BSS, YSS, and RC stars are highly probable members with membership probabilities greater than 0.6 except one RC star (RC3).

We note from the optical CMD that the G-band magnitudes of BSS varies from $\sim$ 0.5 magnitude below MSTO to $\sim$3 magnitudes above the MSTO. In the UV CMD, the BSS sequence stands out as the F148W magnitudes of the BSS vary from $\sim$2 to $\sim$7 magnitudes above the MSTO. We notice a few red giant branch stars brighter than the PARSEC isochrone in the optical CMD, but are not detected in the UV CMD. 

\subsection{Spectral Energy Distributions of BSS}
\label{Section 3.3}

The characterization of the BSS and detection of any hot companion associated with them is accomplished by constructing their spectral energy distributions (SEDs). We first examined the images of the BSS in \textit{Aladin}\footnote{https://aladin.u-strasbg.fr/} to check if there were any nearby (within 3$\arcsec$) sources present. We found BSS4 and BSS6 to have multiple sources within 3$\arcsec$. Moreover, BSS6 is a saturated source. Therefore, these two BSS were excluded from the SED fitting. There is one BSS, BSS9, which is present at the edge of the UVIT image. Its UVIT fluxes may have large errors. However, the GALEX FUV and NUV fluxes are available for this source, and hence we include it in the SED fitting. In order to construct SEDs we made use of virtual observatory SED analyzer (VOSA, \citealt{bayo2008vosa}). Using VOSA, we obtained photometric fluxes of sources in FUV and NUV from GALEX \citep{martin2005galaxy}, optical from \textit{Gaia} EDR3 \citep{2021A&A...649A...1G} and PAN-STARRS \citep{chambers2016pan}, near-IR from Two Micron All Sky Survey (2MASS, \citealt{cohen2003spectral}), and far-IR from Wide-field Infrared Survey Explorer (WISE, \citealt {wright2010wide}). The photometric fluxes were corrected for extinction by VOSA according to the extinction law by \cite{fitzpatrick1999correcting} and \cite{indebetouw2005wavelength} using the value of extinction A$_{v}$ = 0.248 provided by us. We obtained this value of average extinction in the cluster from \citet{knudstrup2020extremely,xin2005blue}. 
The values of fluxes of the BSS in different filters are listed in Table \ref{table: Table 3}. VOSA calculates synthetic photometry for selected theoretical models using filter transmission curves and performs a $\chi^{2}$ minimization test by comparing the synthetic photometry with the extinction corrected observed fluxes to get the best-fit parameters of the SED. The reduced $\chi_{r}^{2}$ is determined using the following formula: 
\begin{equation} 
   \chi_{r}^{2} =\frac{1}{N-N_{f}}\sum_{i=1}^{N} \frac{(F_{o,i}-M_{d}F_{m,i})^{2}}{\sigma^{2}_{o,i}}
\end{equation} 

where N is the number of photometric data points used and N$_{f}$ is the number of free parameters in the model. $F_{o,i}$ and F$_{m,i}$ are the observed and the model flux of the star, respectively. M$_{d}$ is the scaling factor by which the model is to be multiplied to get the fit and is given by $(\frac{R}{D})^{2}$, where R is the radius of the star, D is the distance to the star, and $\sigma_{o,i}$ is the error in the observed flux. 

\begin{figure*}
    \centering
    \begin{subfigure}[b]{0.48\textwidth}
        \includegraphics[width=1.0\textwidth]{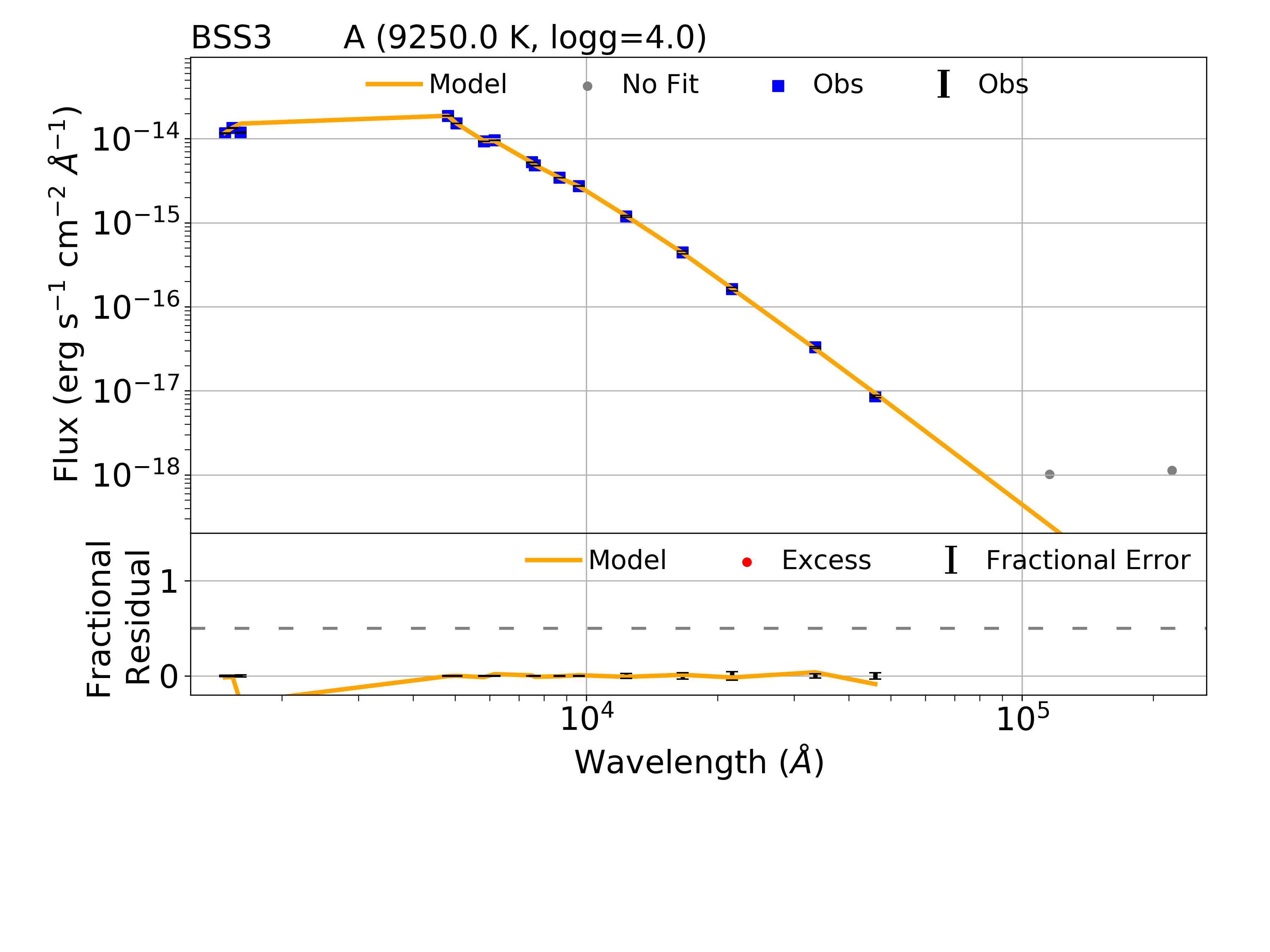}
        \caption*{}
    \end{subfigure}
    \quad
    \begin{subfigure}[b]{0.48\textwidth}
        \includegraphics[width=1.0\textwidth]{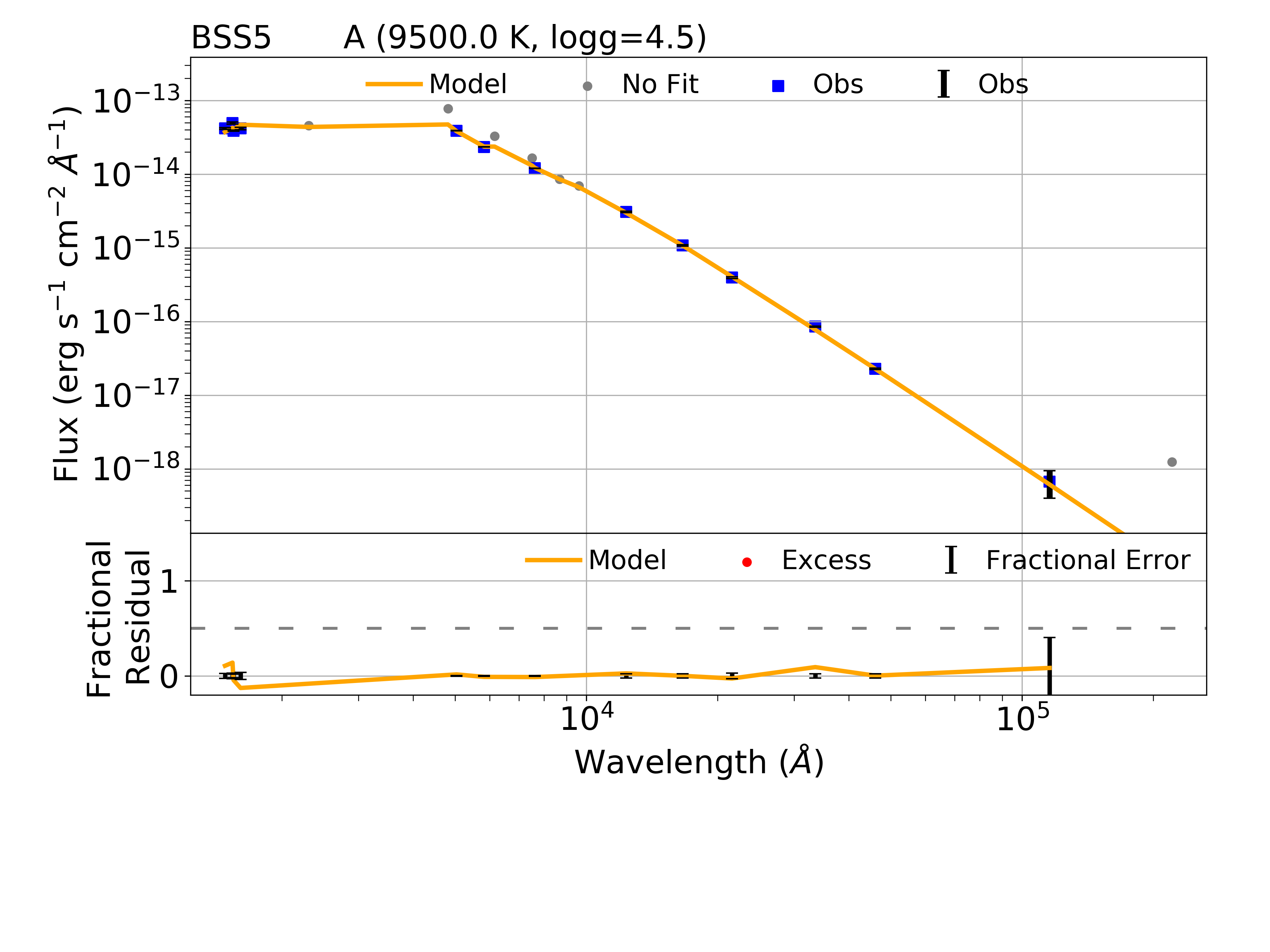}
        \caption*{}
    \end{subfigure}
    \quad
    \begin{subfigure}[b]{0.48\textwidth}
        \includegraphics[width=1.0\textwidth]{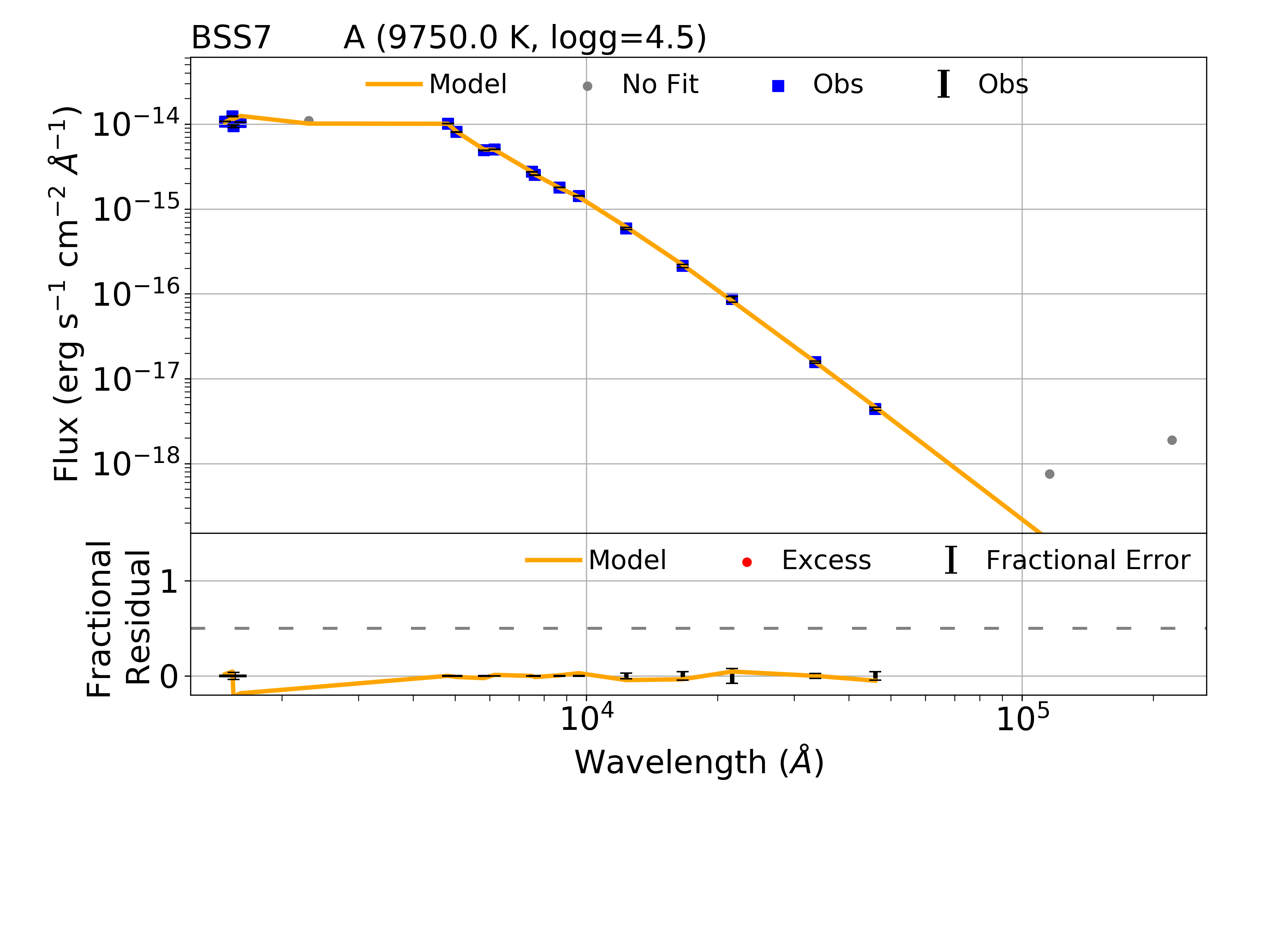}
        \caption*{}
    \end{subfigure}
    \quad
    \begin{subfigure}[b]{0.48\textwidth}
        \includegraphics[width=1.0\textwidth]{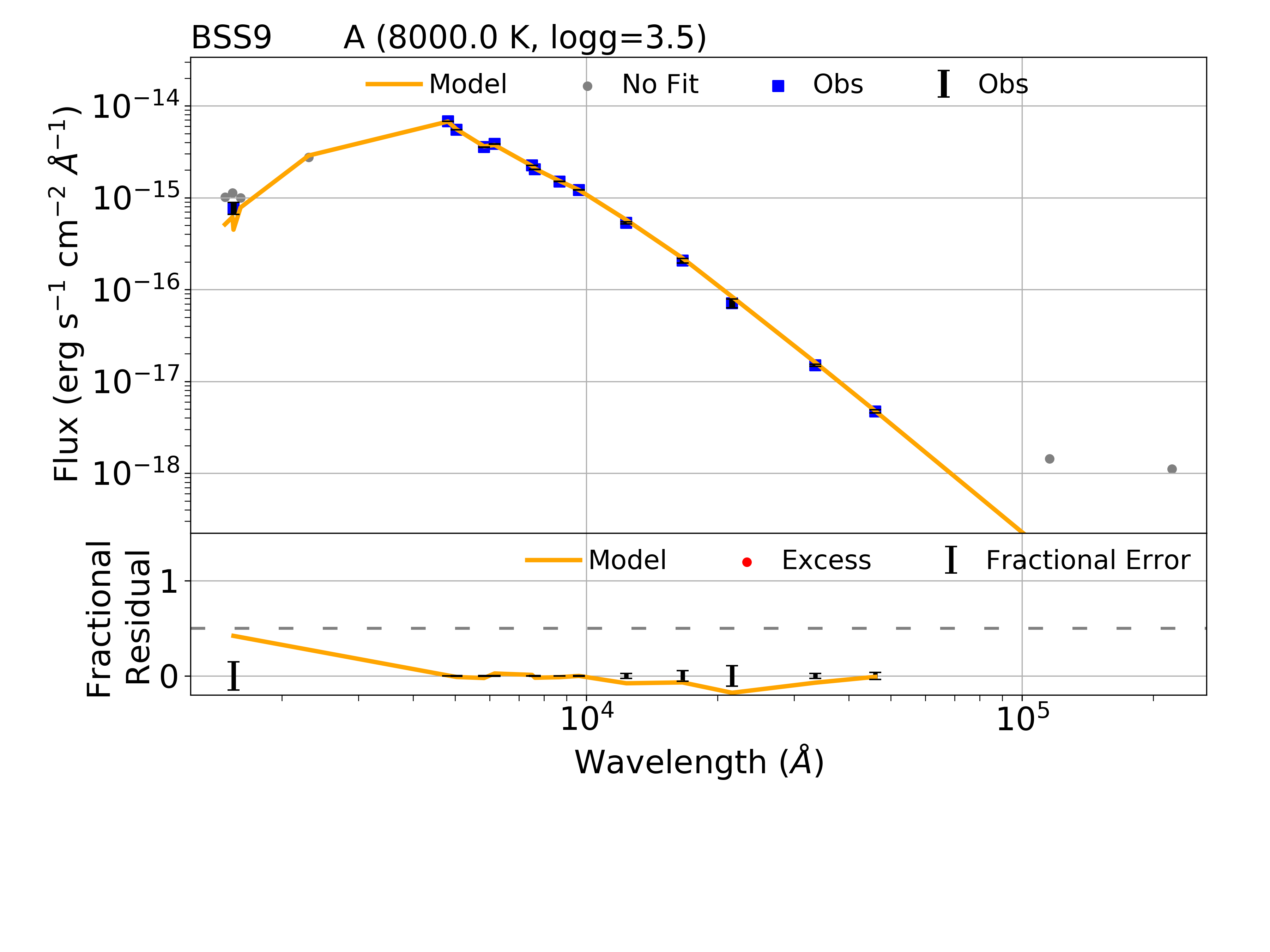}
        \caption*{}
    \end{subfigure} 
\caption{The single component SED fits of BSS. In the top panel, blue data points show the extinction corrected observed fluxes, with black error bars representing the errors in observed fluxes, and the orange curve representing the Kurucz stellar model fit. The bottom panel shows the residual between extinction-corrected observed fluxes and the model fluxes across the filters from UV to IR wavelengths.}
\label{Fig. 3}
\end{figure*}

\begin{figure*}
    \centering
    \begin{subfigure}[b]{0.48\textwidth}
        \includegraphics[width=1.0\textwidth]{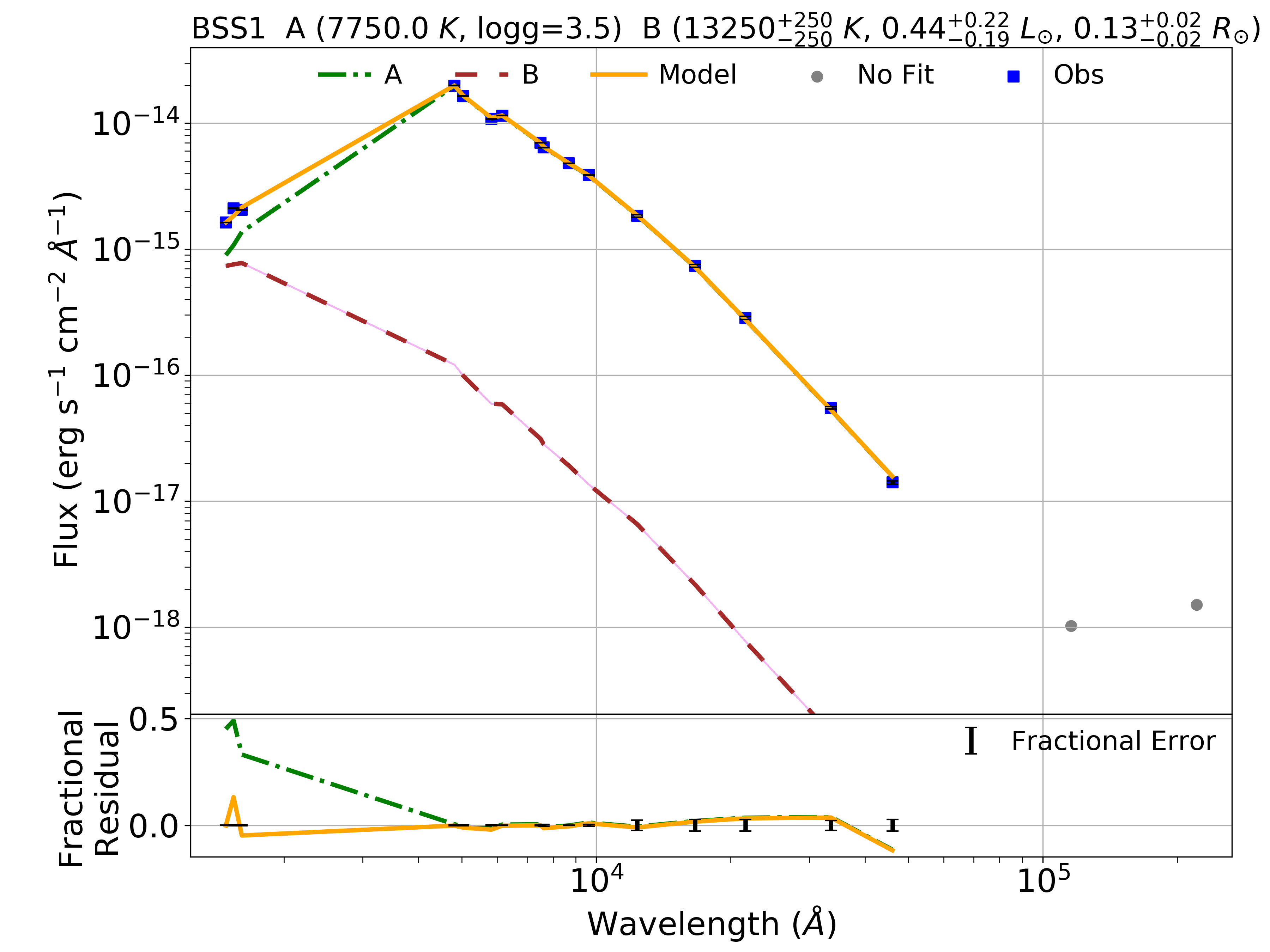}
        \caption*{}
    \end{subfigure}
    \quad
    \begin{subfigure}[b]{0.48\textwidth}
        \includegraphics[width=1.0\textwidth]{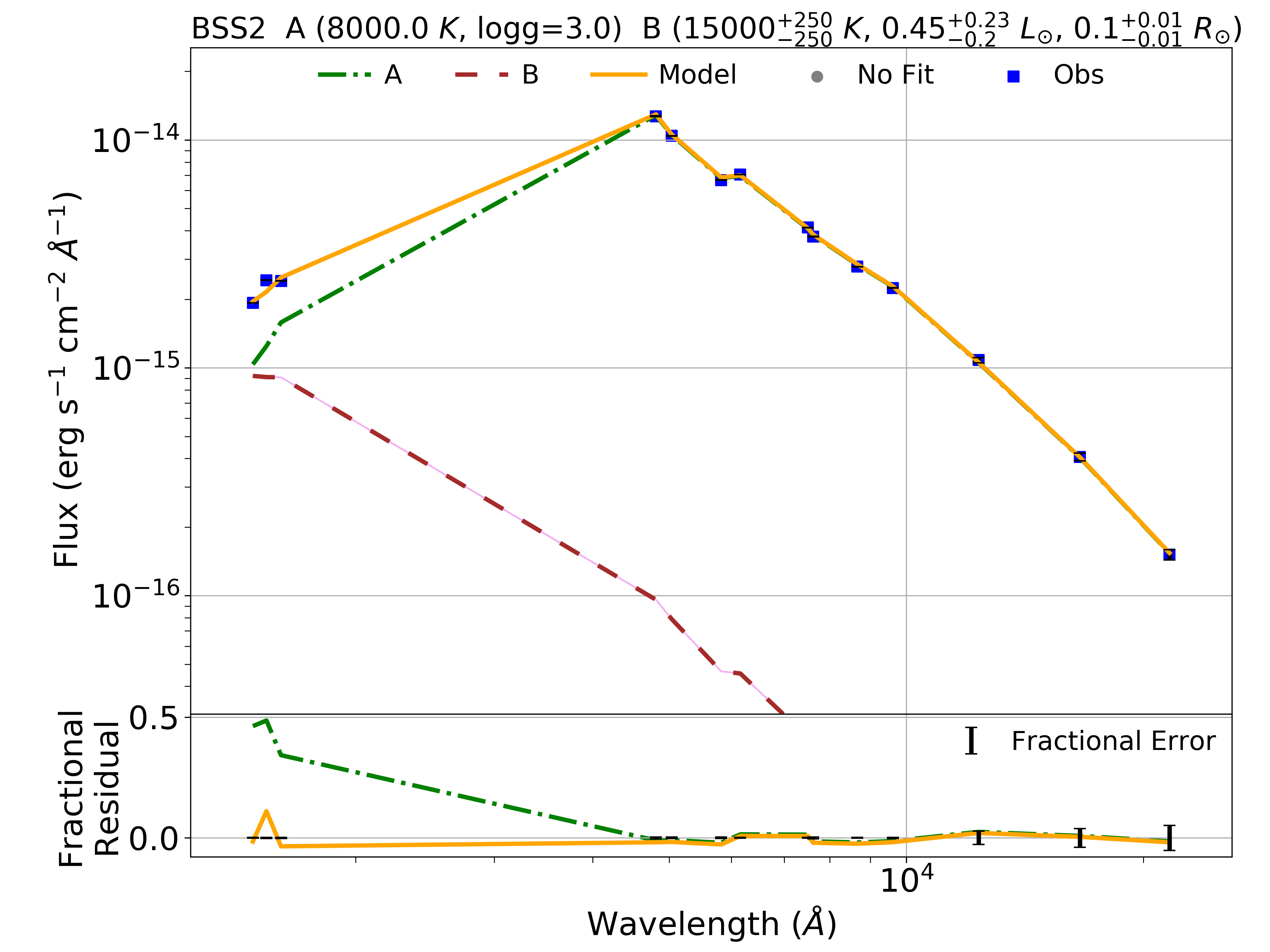}
        \caption*{}
    \end{subfigure}
    \quad
    \begin{subfigure}[b]{0.48\textwidth}
        \includegraphics[width=1.0\textwidth]{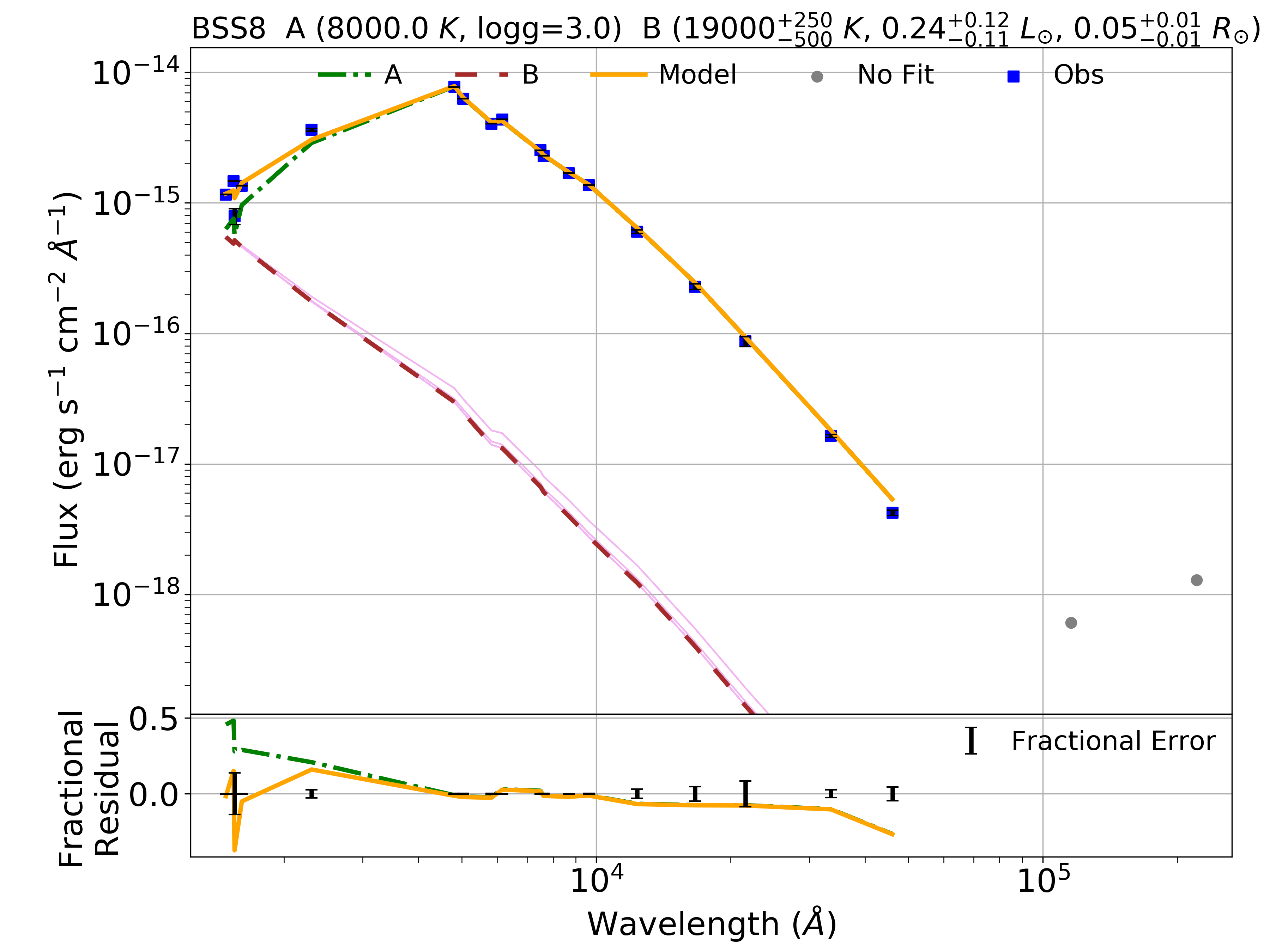}
        \caption*{}
    \end{subfigure}
    \quad
\caption{The binary component fit SEDs of BSS. The top panel shows the double component SED of each BSS where blue data points are the extinction corrected flux values with flux errors as black error bars. The green dashed line is the cool (A) component fit, the brown dashed line is the hot (B) component fit with pink curves denoting the residual of iterations. The composite fit is shown in orange curve and the data points that are not included in the fits are denoted by grey data points. The bottom panel shows the fractional residual for both single (green) and composite (orange) fits. The fractional errors are shown on the x-axis by black error bars. The parameters of the cool and hot components which were derived from the SEDs along with the estimated errors are mentioned at the top of the plots.}
\label{Fig. 4}
\end{figure*}

\begin{figure*}
    \centering
    \begin{subfigure}[b]{0.48\textwidth}
        \includegraphics[width=1.0\textwidth]{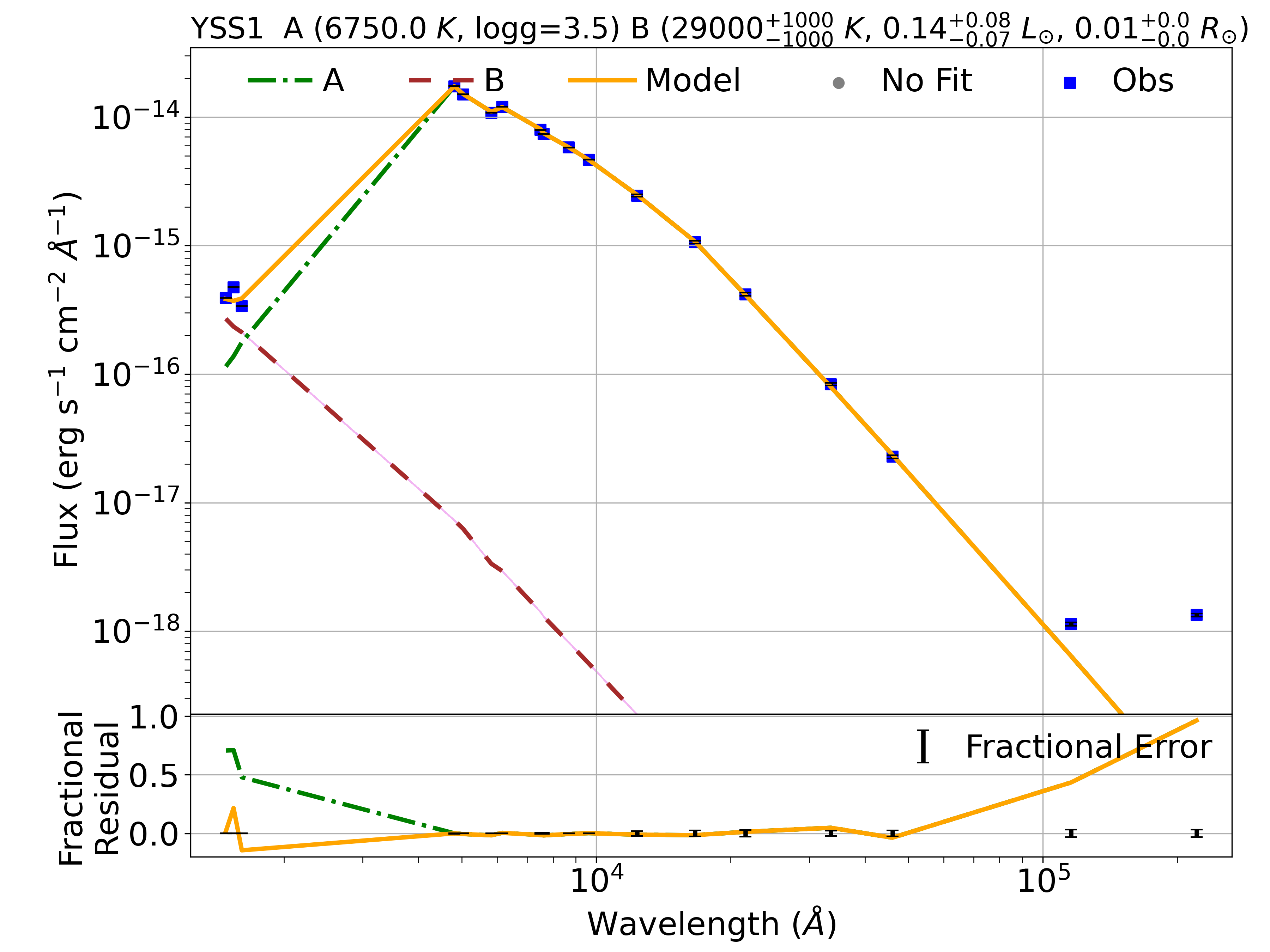}
        \caption*{}
    \end{subfigure}
    \quad
    \begin{subfigure}[b]{0.48\textwidth}
        \includegraphics[width=1.0\textwidth]{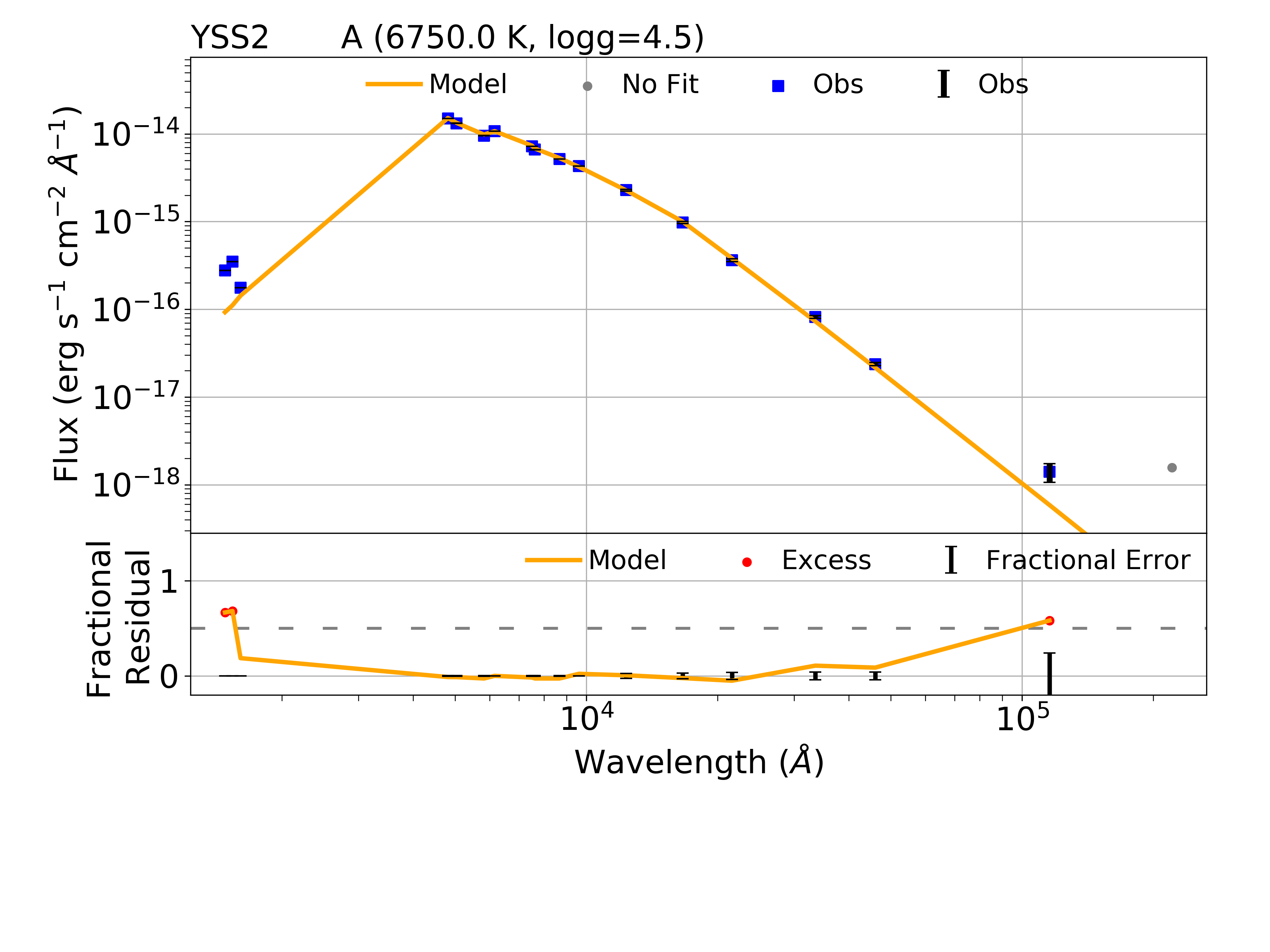}
        \caption*{}
    \end{subfigure}
    \quad
    \begin{subfigure}[b]{0.48\textwidth}
        \includegraphics[width=1.0\textwidth]{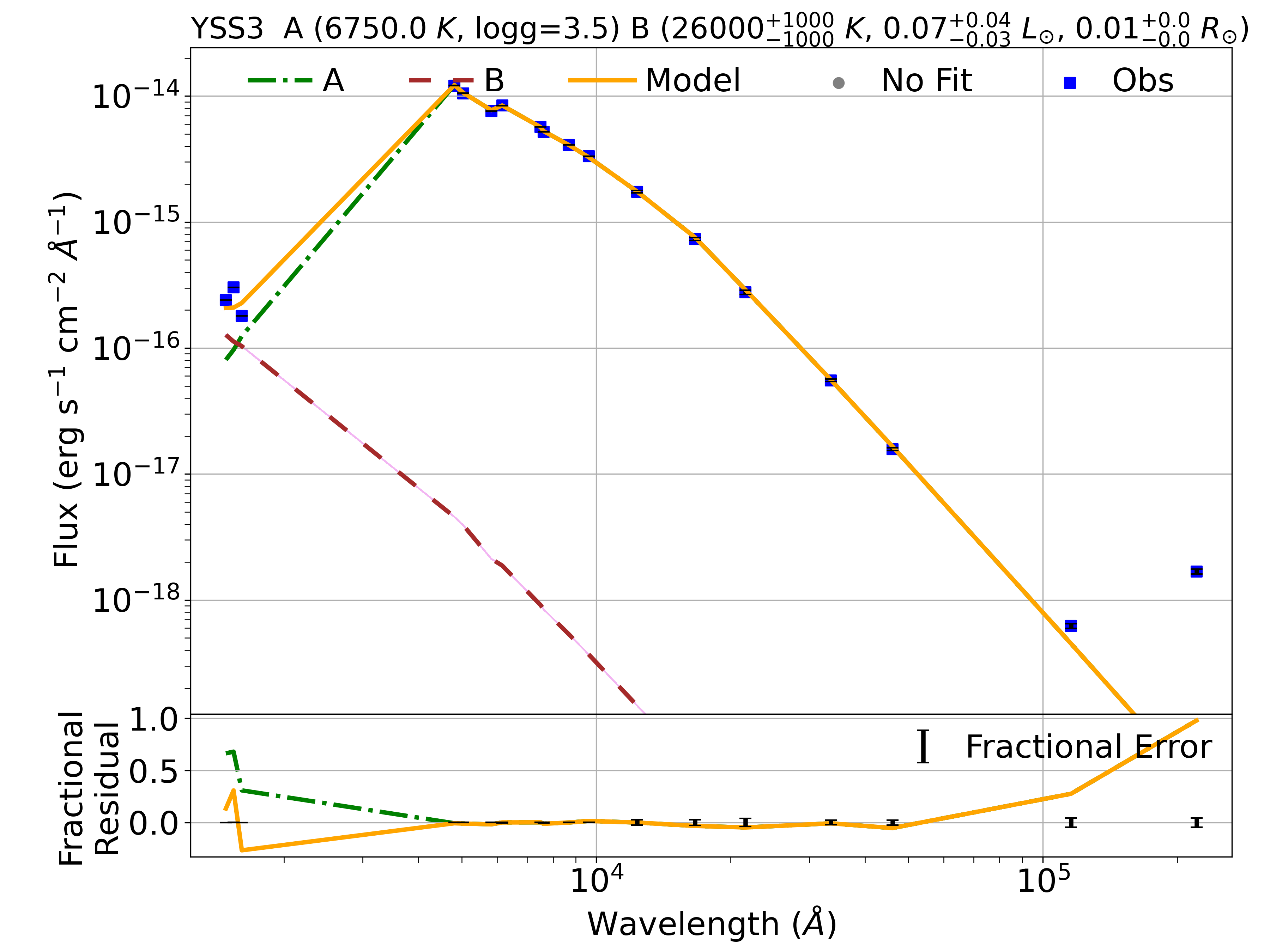}
        \caption*{}
    \end{subfigure}
    \quad
    \begin{subfigure}[b]{0.48\textwidth}
        \includegraphics[width=1.0\textwidth]{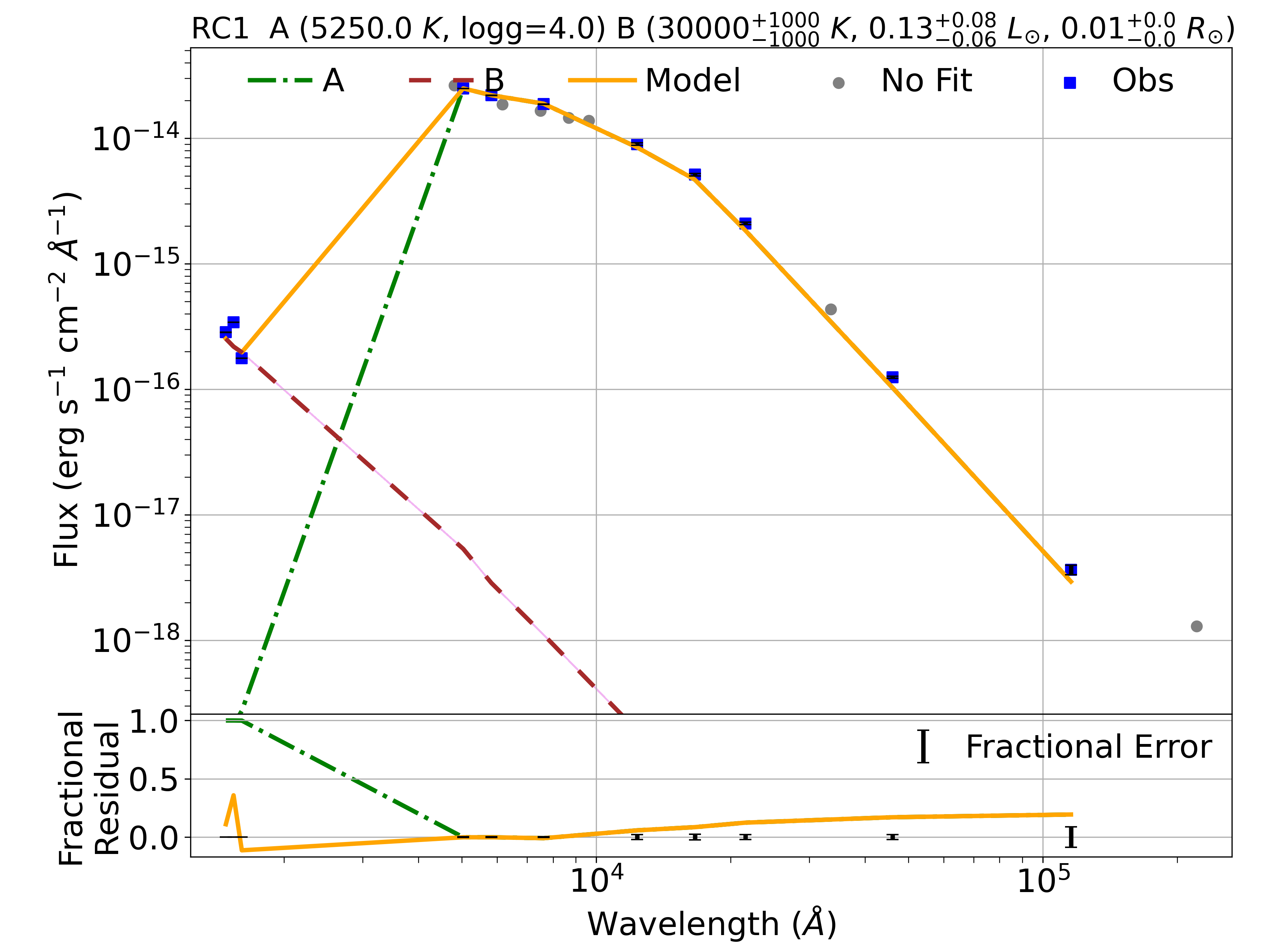}
        \caption*{}
    \end{subfigure}
    \quad
    \begin{subfigure}[b]{0.48\textwidth}
        \includegraphics[width=1.0\textwidth]{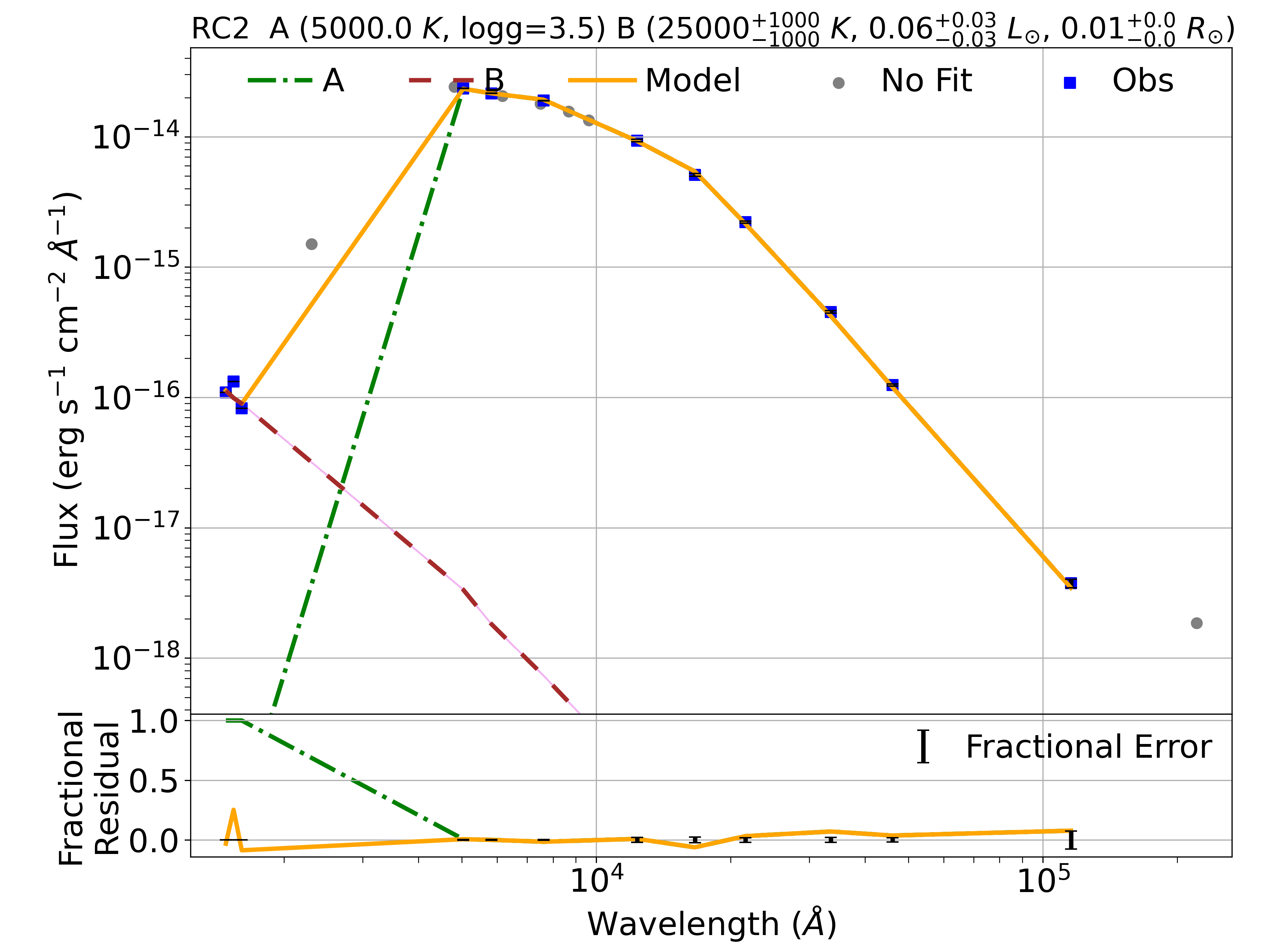}
        \caption*{}
    \end{subfigure}
    \quad
    \begin{subfigure}[b]{0.48\textwidth}
        \includegraphics[width=1.0\textwidth]{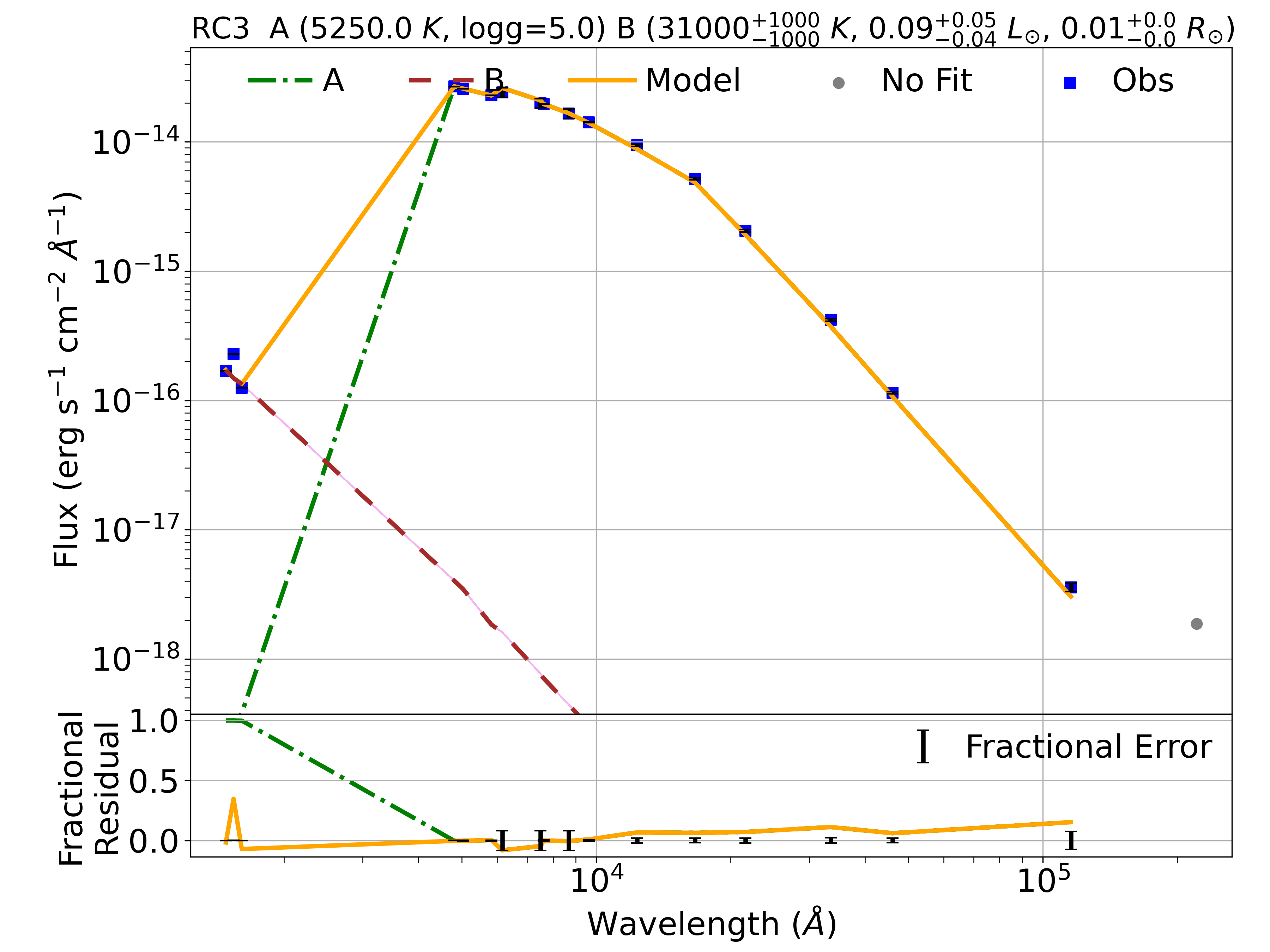}
        \caption*{}
    \end{subfigure}
\caption{SEDs of YSS and RC stars. The symbols and curves mean the same as in Figure \ref{Fig. 3}.}
\label{Fig. 5}
\end{figure*}

In order to fit the SEDs to the BSS, we used the Kurucz stellar models \citep{castelli1997notes}. We kept T$\mathrm{_{eff}}$ and log g as free parameters and chose their ranges to be 3500--50000 K and 3--5, respectively. We fixed the value of metallicity ([Fe/H]) to be $-$0.5 which is nearest to the cluster metallicity $-$0.36 \citep{knudstrup2020extremely}. While fitting the SEDs, we first excluded the UV data-points from the SEDs and confirmed whether optical and IR data points were fitting satisfactorily with the model flux. We checked carefully if excess to the UV and/or IR data points were present and also noted the residuals (the difference between model flux and the observed flux) in all the UV filters. Out of 7 BSS, 4 showed fractional residual less than 0.3 in all the UV data points, and therefore we fitted them with single component SEDs. Figure \ref{Fig. 3} shows all the single component SEDs. The top panel for each BSS shows the fitted SED of each BSS where blue data points are the extinction corrected observed fluxes with error bars shown in black, and the orange curve is the model fit. The bottom panel shows the residual between observed fluxes and the model fluxes in each filter. In BSS5, BSS7, and BSS9, fluxes were available in both GALEX/FUV and GALEX/NUV filters. However, GALEX/NUV points were flagged as bad in the GALEX catalogue \citep{bianchi2000ultraviolet} itself, hence we did not use them in the fitting. As mentioned above, since BSS9 is located at the edge of the UVIT image, UVIT data points of this BSS are excluded from the SED fitting. In BSS3, BSS7, and BSS9, the upper limits were available for both WISE W3 and W4, whereas in the case of the BSS5, the upper limit was available in the WISE W4 filter. Hence, these data points are excluded from the corresponding SEDs of BSS. We note that for these BSS (BSS3, BSS5, BSS7, and BSS9), the residuals are coming out to be nearly zero, indicating that the extinction corrected observed fluxes and the model fluxes in all filters are comparable. The satisfactory fitting of the single component SED suggests that there are no signatures of the presence of any hotter companions associated with these four BSS.

The parameters of all the BSS obtained from double component SEDs, are listed in Table \ref{table4}. The $\chi^{2}$ of the fits are large even when the SED fits are visually good due to some data points with very small observational flux errors \citep{rebassa2021white}. In view of this fact, VOSA also determines another parameter called visual goodness of fit (vgf$_{b}$). It is a modified reduced $\chi^{2}$ which is calculated by forcing the observational errors to be at least 10$\%$ of the observed flux. It is determined using the following formula:
\begin{equation}
  vgf_{b} =\frac{1}{N-N_{f}}\sum_{i=1}^{N} \frac{(F_{o,i}-M_{d}F_{m,i})^{2}}{b^{2}_{i}} \\
\end{equation}

where, $\sigma_{o,i}\leq {0.1}  F_{o,i} \implies b_{i} = 0.1  F_{o,i}$ and $\sigma_{o,i}\geq{0.1  F_{o,i}} \implies b_{i} = \sigma_{o,i}$. We note that the values of vgf$_{b}$ parameters are < 2 for all single and double component SEDs. These fits are acceptable since the value of vgf$_{b}$ < 15 is an indicative of good SED fits \citep{jimenez2018white,rebassa2021white}.

Three BSS (BSS1, BSS2, and BSS8) show fractional residual greater than 0.3 in the UV data points. Therefore, these 3 BSS are fitted with a double component using a python code, \textsc{Binary SED Fitting}\footnote{https://github.com/jikrant3/Binary SED Fitting}, by \cite{jadhav2021uocs}  which is based on $\chi^{2}_{r}$ minimization technique to fit the double component SEDs. To fit the hotter component, we used the Koester model \citep{koester2010white}, since this model gives a temperature range of 5000--80000 K and log g range of 6.5--9.5. Figure \ref{Fig. 4} shows the double component SEDs for BSS1, BSS2, and BSS8. For each BSS, the top panel shows the fitted SED, and the bottom panel shows the residual for single and composite fit in every filter.
The composite fit satisfactorily takes care of the excess in the UV data points and the residual turns out to be nearly zero in all the data points, which is reflecting in the significantly lower $\chi^{2}_{r}$ values of the double fit as listed in Table \ref{table4}.  
For these BSS, the parameters of the cooler components are taken from VOSA, whereas the parameters of the hotter companions are taken from \textsc{Binary SED Fitting}, by \cite{jadhav2021uocs}. 
In order to determine the errors in the parameters of hot companions, we have followed the statistical approach as described in \cite{2021JApA...42...89J}. For this, we generated 100 iterations of observed SEDs for each BSS by adding Gaussian noise, proportional to the errors, to each data point. 
We fitted these 100 SEDs using the Koester model as described earlier, and derived the parameters of the hotter companions based on these SED fittings.
We considered the median values of the parameters derived from the 100 SEDs to be the parameters of hot companions, whereas, for the errors in the parameters, we considered the standard deviation from the median parameters. If the statistical error is less than the step size of stellar models, half of the step size (e.g. 250 K for Kurucz model) is taken as temperature error.

\begin{figure*}
\includegraphics[width=15cm]{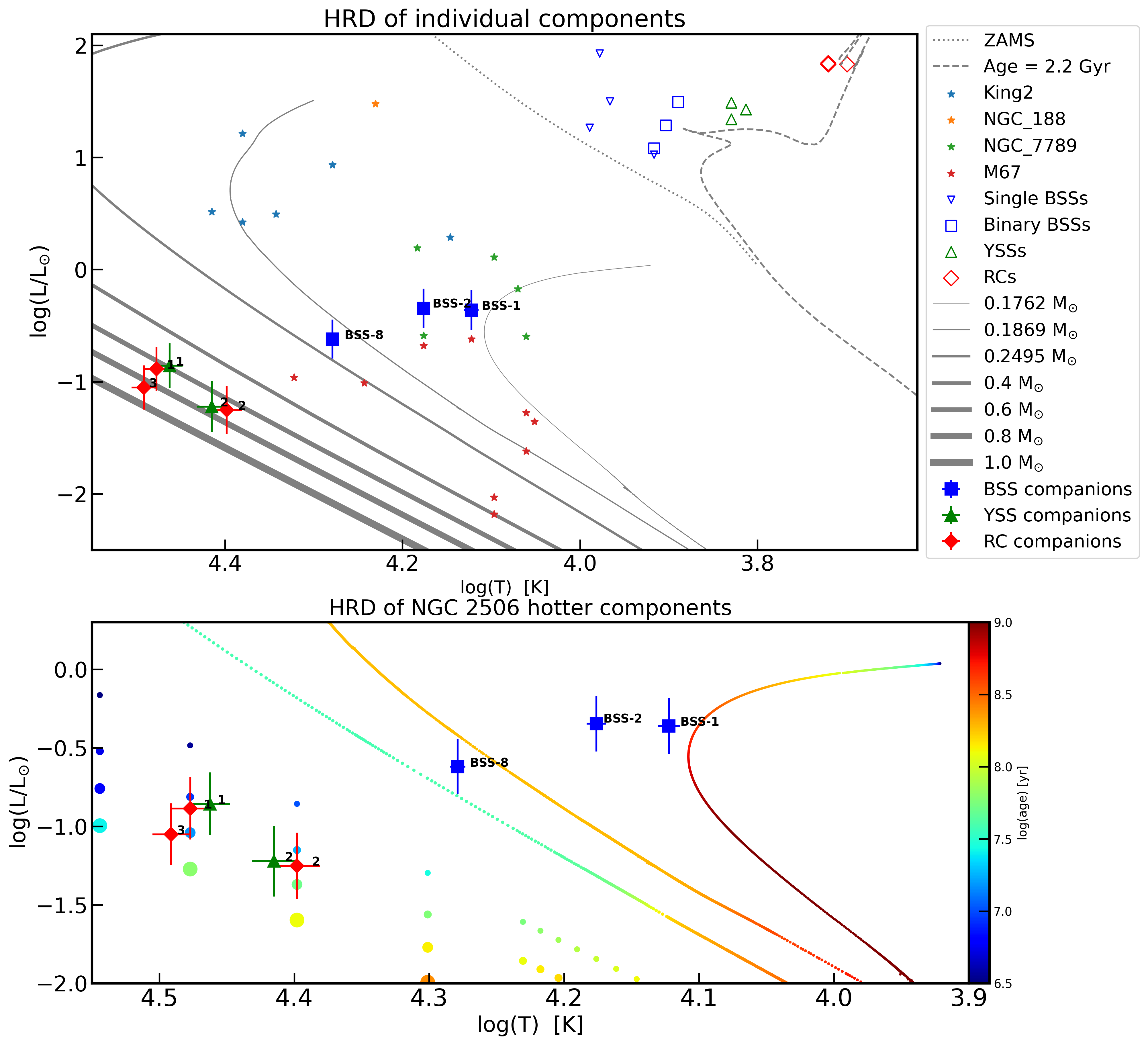}

\caption{The upper panel represents the H-R diagram showing the single component BSS as blue open triangles, the cooler component BSS as blue open squares, and their corresponding hotter companions as blue filled squares. Hotter companions of RC stars and YSS are denoted by red diamonds and green triangles respectively. Their corresponding cooler components are denoted by open diamonds and triangles of the same colours. Hotter companions of BSS of other OCs are shown in different symbols. A PARSEC isochrone of age 2.2 Gyr is plotted as the grey dotted curve and ZAMS is plotted as the grey dashed curve. LM and ELM WD cooling curves of different masses taken from \citet{panei2007full} and \citet{althaus2013new} respectively are represented by different solid grey curves. The lower panel shows the hot companions of NGC 2506 lying on the LM and ELM WD cooling curves of different masses taken from \citet{panei2007full} and \citet{althaus2013new} respectively indicating their approximate cooling ages.}

\label{Fig. 6}
\end{figure*}
 
\subsection{Spectral Energy Distributions of YSS and RC stars}
We also constructed the SEDs for the 3 YSS and 3 RC candidates detected in all UVIT/FUV filters using VOSA, following the same steps as described for BSS using Kurucz model \citep{castelli1997notes}. All these 6 sources showed excess in UV data points, suggesting the presence of hot companions. In all the cases, composite fit has reduced $\chi^{2}$ of the fit significantly compared to single fits, except YSS2 for which no satisfactory double fit was found. Therefore, only the single component SED has been shown for YSS2. The SEDs of YSS and RC stars are shown in Figure \ref{Fig. 5} and their estimated parameters are tabulated in Table \ref{table4}. We also fitted the primary component of our objects (BSS, YSS, and RC stars) with Coelho synthetic stellar library \citep{coelho2014new}. We found that sources with UV excess in the Kurucz stellar models also showed similar UV excess when fitted with the Coelho model. This implies that the residuals in the UV data points are independent of the stellar models used.
However, in this work we have presented the results of SED fits using Kurucz model.

\section{Discussions} \label{Section 4}

\subsection{BSS, YSS, and RC stars properties}
As mentioned above, on fitting the SEDs, we get the parameters such as temperatures, radii, and luminosities of BSS, YSS, and RC stars as well as their hotter components. 
The hot temperatures (7750--9750 K) of NGC 2506 BSS are consistent with the young age (2.2 Gyr) of the cluster. This temperature range of BSS is comparable to temperatures of the BSS of other intermediate-age OCs such as NGC 7789 that varies from 7250--10250 K \citep{vaidya2022uocs} and NGC M67 having a temperature range of 6250--9000 K \citep{sindhu2019uvit,pandey2021uocs}.
We note that from the SEDs, that the temperatures of YSS varies from 6500--6750 K. This shows that YSS are cooler than BSS which is expected since the YSS are presumably  evolved BSS. Moreover, RC stars have temperature that varies from 5000--5250 K. We compared the SED-based temperatures of our objects (BSS, YSS, and RCs) with the BP/RP spectra-based temperatures from Gaia DR3 \citep{de2022gaia,babusiaux2022gaia}. Our SED-estimated temperatures match Gaia DR3 temperatures to within 400 K for objects that have single-temperature fits, but vary greatly in objects found with hot companions. This is not surprising as the spectrophotometric parameters are not likely to be accurate in case of binaries.

We made a comparison with ZAMS and found that all BSS have masses of 1.61--2.16 M$_\odot$. On comparing with the turnoff mass, 1.45 M$_\odot$ of the cluster, we suggest that these BSS must have at least gained $\sim$ 0.16--0.71 M$_\odot$ through the mass transfer process. 

We checked for the information on variability of BSS, YSS, and RC stars in Gaia DR3. Two of our objects, BSS1 and YSS1 are found to be short timescale (< 0.5 -1 day) main sequence type oscillators (GDOR|DSCTU|SXPHE).

\subsection{The nature of hot companions}
We plot the H-R diagram as shown in Figure \ref{Fig. 6} to understand the nature of the hot companions of the BSS, YSS, and RC stars. The upper panel of this diagram includes a PARSEC isochrone of age = 2.2 Gyr, distance = 3110 pc, Z = 0.0045, and a zero age main sequence (ZAMS). 
We have shown the cooler as well as hotter companions of BSS, YSS, and RCs of NGC 2506 along with the hot companions of BSS of other OCs. We notice that the single SED fitted BSS are bluer than the A-component of the double temperature fitted BSS. As mentioned earlier, the YSS (both single and double SED fitted) are brighter than the sub-giant branch and cooler than the BSS. It is interesting to note three sequences of single BSS, binary BSS, and YSS in the H-R diagram from hotter to cooler temperatures. RC1-A, RC2-A, and RC3-A lie on the location of the RC stars in the H-R diagram. 

Among the hot companions of BSS, BSS8-B lies on the WD cooling curve of mass 0.20 M$_\odot$ \citep{panei2007full} suggesting it to be a low mass (LM) WD. The locations of BSS1-B and BSS2-B in the H-R diagram suggest them to be extremely low mass (ELM) WDs since they are lying very close to the ELM WD cooling curves \citep{althaus2013new} of masses 0.17 M$_\odot$ and 0.18 M$_\odot$ respectively. ELM WDs are the stellar remnants that do not ignite helium in their cores \citep{brown2010elm}. The detection of LM/ELM WD companion supports the Case A/Case B mass transfer formation mechanism of BSS because the LM WDs (masses < 0.4 M$_\odot$) and ELM WDs (masses < 0.2 M$_\odot$) cannot form from a single star evolution within Hubble time \citep{brown2011binary}. This implies that these WDs must have undergone mass loss during their evolution. The location of NGC 2506 BSS hot companions in the H-R diagram is similar to the hot companions of BSS in other intermediate-age OCs such as M67 \citep{sindhu2019uvit} and NGC 7789 \citep{vaidya2022uocs} which also show LM/ELM WDs candidates for some of their BSS. 

The hot companions of RC1 and RC2 lie on the Bergeron WD model of mass $\sim$ 0.6 M$_\odot$, whereas that of RC3 lies on the Bergeron WD model of mass $\sim$ 0.8 M$_\odot$. It can be noted that the hot companion of YSS1 and YSS3 also lie on the Bergeron WD model of masses $\sim$ 0.6 M$_\odot$ and $\sim$ 0.8 M$_\odot$ respectively. The hot companions of the 2 YSS and 3 RC stars are hotter and less luminous than the above hot companions of BSS.
The lower panel of Figure \ref{Fig. 6} shows the approximate ages of the hot companions of BSS according to \citet{panei2007full} and \citet{althaus2013new} models and that of RC stars and YSS according to the Bergeron WD models \citep{tremblay2009spectroscopic}. It can be inferred that the hot companions of BSS have log age of $\sim$ 7.5--8.5, hot companions of YSS have log age of $\sim$ 7.0--8.0, and log age of hot companions of RC stars varies from $\sim$ 6.5--7.5. 

It is noteworthy that radial velocity (RV) for these three red clump sources are available in \textit{Gaia} DR2. Their RVs are found as RC1 = 79.98 $\pm$ 2.13 km s$^{-1}$, RC2 = 93.79 $\pm$ 1.60 km s$^{-1}$, and RC3 = 82.19 $\pm$ 1.73 km s$^{-1}$. These values are consistent with the cluster mean RV, 84 km s$^{-1}$, and therefore these RCs should indeed be the members of the cluster. 
Moreover, RC1 is a probable binary according to the spectroscopic study by \citet{anthony2018wiyn} where they have reported its RV to be 82.1 km s$^{-1}$. 

In the absence of comprehensive RV studies, for most cluster sources, their spectroscopic binary fraction are not known. The binarity information of our sources (BSS, YSS, and RC stars) is not available in Gaia DR3 \citep{eyer2022gaia}. From our SED analysis, we find that at least 3 out of 7 (42$\%$) BSS in this cluster are formed through the mass transfer. We also find 2 out 3 YSS in this cluster to be binaries. Among the BSS+YSS population, the binary fraction is 50\%, suggesting a significant pathway of formation through mass transfer. The above estimations provide a lower limit, as they are based on systems with detectable hot companions. Recently, a 1.6 Gyr old cluster NGC 7789, was found to have at least 33\% of the BSS to have formed via mass transfer \citep{vaidya2020blue}.
 
The detected hot companions to the BSS/YSS/RCs show a large spread in mass and age. The mass  range covers low to high mass WDs (0.2--0.8 M$_\odot$), whereas the turn-off star of this cluster will evolve to a $\sim$ 0.6 M$_\odot$ WD. The presence of LM WDs suggest the existence of Case A/B mass transfer among close binary systems. The three systems (2 RCs and 1 YSS) with a normal mass ($\sim$ 0.6 M$_\odot$) WD can be formed through the evolution of a normal turn-off star of mass 1.45 M$_\odot$. In these systems, mass transfer may or may not have happened as it will depend on the binary period. However, the presence of high mass (> 0.6 M$_\odot$) WDs in two systems (YSS3 \& RC3) suggests that the progenitors were massive than turn-off stars, likely BSS, demanding that the progenitors need to be higher order systems. We detect 2 such systems (YSS3 \& RC3) which should have been at least triples. \citet{knudstrup2020extremely} studied three detached eclipsing binaries in this cluster. They classified one among these three to be a likely triple system which is also detected in the UVIT/FUV images. Their detailed analysis suggested an inner close binary (1.47 \& 1.25 M$_\odot$) of $\sim$ 2.9 d period with a likely $\sim$ 0.7  M$_\odot$ star in an eccentric large ($\sim$ 443 d) orbit, that could either be a WD or a main-sequence star. Therefore, at least one triple system with close inner binary is present in the cluster. It is quite possible that similar systems could be the progenitors of YSS3 and RC3. These two systems suggest that the BSS formation pathway through triple systems are indeed operational in this cluster. More importantly, close binaries, either in binaries or triples are indeed important for BSS formation pathway in OCs.

\section{Summary} \label{Section 5}
The work presented in this paper can be summarized as follows:

\begin{enumerate}

\item We identified 2175 cluster members in NGC 2506 using a machine learning based algorithm, ML--MOC, on Gaia EDR3 data including 9 BSS, 3 YSS, and 3 RC stars. We present the analysis of these BSS, YSS, and RC stars using \textit{ASTROSAT}/UVIT data in three FUV filters, F148W, F154W, and F169M.
    
\item The multi-wavelength SEDs of only 7 BSS were constructed as two of them had neighbours within 3$\arcsec$. Out of 7 BSS, we found that 4 BSS fitted well with a single temperature SED, whereas 3 showed an excess greater than 30$\%$ from the best-fit model. These 3 BSS were fitted with double component SEDs and the properties of the hot companions are reported in this work.

\item The temperatures of BSS of NGC 2506 varies from 7750 -- 9750 K which are consistent with the young age of the cluster. The temperatures of YSS varies from 6500 -- 6750 K, whereas the temperatures of RC stars varies from 5000 -- 5250 K.
 
\item We discover two ELM WD and one LM WD as companions to three BSS in the cluster. One of the ELM WD of mass $\sim$0.18 M$_\odot$ is found as a companion to BSS1. Its temperature is estimated to be 13250 K, luminosity to be 0.44 L$_\odot$, and radius to be 0.13 R$_\odot$. The second ELM also of mass $\sim$0.18 M$_\odot$ is found as a companion of BSS2. It has temperature $\sim$ 15000 K, luminosity $\sim$ 0.45 L$_\odot$, and radius $\sim$ 0.10 R$_\odot$. The LM WD of mass $\sim$0.20 M$_\odot$ is found as a companion to BSS8 with T$\mathrm{_{eff}}$ $\sim$19000 K, L $\sim$0.24 L$_\odot$, and R $\sim$ 0.05 R$_\odot$. The hot companions of these BSS have log age $\sim$7.5 -- 8.5. 
  
\item We also constructed the SEDs of the 3 YSS and 3 RC candidates that were detected in UVIT/FUV filters. All these 6 objects showed an excess in FUV data points indicating the presence of a hotter companion. We fitted the double component SEDs of all RCs and YSS (except YSS2). From the parameters (T$\mathrm{_{eff}}$ $\sim$ 29000 -- 31000 K, L $\sim$ 0.05 -- 0.14 L$_\odot$, and R $\sim$ 0.01 R$_\odot$) of the hot companions of YSS1, RC1, and RC2, we infer that they are likely to be normal mass ($\sim$ 0.6 M$_\odot$) WDs, suggesting to be formed from the star of mass 1.45 M$_\odot$ which is the cluster turn-off mass. However, the presence of high mass ($\sim$ 0.8 M$_\odot$) WD with parameters T$\mathrm{_{eff}}$ $\sim$ 26000 -- 31000 K, L $\sim$ 0.07 -- 0.09 L$_\odot$, and R $\sim$ 0.01 as the companions of YSS3 and RC3 indicates the presence of their massive progenitors such as BSS. Thus, these two systems may likely have formed from triplets.  

\item We conclude that in the open cluster NGC 2506, the Case A/Case B mass transfer mechanism is likely to be responsible for the formation of at least 4 out of 10 (40\%) BSS and YSS systems. However, merger in triple system with close inner binary is the potential formation pathway of YSS and RC stars with BSS as their progenitors.
\end{enumerate}

\section*{Acknowledgements}

We thank anonymous referee for the valuable comments. AP and KV acknowledge the financial support from Indian Space Research Organization (ISRO) under the AstroSat archival data utilization program (No. DS 2B-13013(2)/3/2019-Sec.2). AS acknowledges the support from SERB POWER fellowship grant SPF/2020/000009. This work uses the data from UVIT onboard AstroSat mission of Indian Space Research Organisation (ISRO). UVIT is a collaborative project between Indian Institute of Astrophysics (IIA), Bengaluru, The Indian-University Centre for Astronomy and Astrophysics (IUCAA), Pune, Tata Institute of Fundamental Research (TIFR), Mumbai, several centres of Indian Space Research Organisation (ISRO), and Canadian Space Agency (CSA). This work has made use of TOPCAT \citep{taylor2011topcat}, Matplotlib \citep{hunter2007matplotlib}, IPython \citep{perez2007ipython}, Scipy \citep{oliphant2007scipy,millman2011python} and Astropy, a community-developed core Python package for Astronomy \citep{price2018astropy}. This publication also makes use of VOSA, developed under the Spanish Virtual Observatory project supported by the Spanish MINECO through grant AyA2017-84089.

\section*{Data availability}

The data underlying this article are publicly available at \url{https://astrobrowse.issdc.gov.in/astro_archive/archive/Home.jsp}
The derived data generated in this research will be shared on reasonable request to the corresponding author.

\bibliographystyle{mnras}
\bibliography{references}

\begin{table*}
\centering
\caption{\label{table: Table 3} Coordinates of all BSS, YSS, and RC stars in Columns 2 and 3, \textit{Gaia} EDR3 ID in Column 4, UVIT F148W and F154W fluxes in Columns 5 and 6, GALEX FUV flux in Column 7, UVIT F169M flux in Column 8, GALEX NUV flux in Column 9, \textit{Gaia} EDR3 Gbp, G, and Grp fluxes in Columns 10--12, 2MASS J, H, and Ks fluxes in Columns 13--15, and WISE W1, W2, W3, and W4 fluxes in Columns 16--19. All flux values are listed in the unit of erg s$^{-1}$ cm$^{-2}$\AA$^{-1}$.}

\begin{tabular}{cccccccccccc}
\hline
\\
Name&RA&DEC&Gaia EDR3 ID&UVIT.F148W$\pm$err&UVIT.F154W$\pm$err&\\
GALEX.FUV$\pm$err&UVIT.F169M$\pm$err&GALEX.NUV$\pm$err&PS1.g$\pm$err&GAIA3.Gbp$\pm$err&GAIA3.G$\pm$err&\\
PS1.r$\pm$err&PS1.i$\pm$err&GAIA3.Grp$\pm$err&PS1.z$\pm$err&PS1.y$\pm$err&2MASS.J$\pm$err&\\
2MASS.H$\pm$err&2MASS.Ks$\pm$err&WISE.W1$\pm$err&WISE.W2$\pm$err&WISE.W3$\pm$err&WISE.W4$\pm$err\\
\\
\hline
\hline
\\
\\
BSS1&119.99207&-10.76498&3038044880608396672&1.630e-15$\pm$8.654e-19&2.110e-15$\pm$4.692e-18&\\
-&2.050e-15$\pm$5.585e-18&-&1.988e-14$\pm$4.604e-17&1.636e-14$\pm$5.059e-17&1.079e-14$\pm$2.799e-17&\\
1.150e-14$\pm$1.021e-17&6.975e-15$\pm$1.596e-17&6.416e-15$\pm$2.300e-17&4.800e-15$\pm$3.276e-18&3.880e-15$\pm$1.548e-17&1.837e-15$\pm$4.398e-17&\\
7.371e-16$\pm$2.037e-17&2.848e-16$\pm$7.871e-18&5.504e-17$\pm$1.267e-18&1.407e-17$\pm$3.759e-19&-&-\\
\hline
\\
BSS2&119.99902&-10.75405&3038046358077125248&1.927e-15$\pm$2.721e-18&2.424e-15$\pm$6.079e-18&\\
-&2.406e-15$\pm$6.101e-18&-&1.272e-14$\pm$5.119e-17&1.042e-14$\pm$2.818e-17&6.667e-15$\pm$1.729e-17&\\
7.054e-15$\pm$8.083e-18&4.129e-15$\pm$5.099e-18&3.769e-15$\pm$1.344e-17&2.780e-15$\pm$3.687e-18&2.242e-15$\pm$5.599e-18&1.081e-15$\pm$2.988e-17&\\
4.065e-16$\pm$1.573e-17&1.512e-16$\pm$7.796e-18&-&-&-&-\\
\hline
\\
BSS3&119.97925&-10.79276&3038044674450004736&1.159e-14$\pm$7.939e-17&1.339e-14$\pm$1.143e-16	&\\
-&1.184e-14$\pm$1.321e-16&-&1.868e-14$\pm$6.800e-17&1.518e-14$\pm$4.195e-17&9.267e-15$\pm$2.393e-17&\\
9.544e-15$\pm$3.118e-17&5.226e-15$\pm$6.397e-18&4.800e-15$\pm$1.740e-17&3.433e-15$\pm$1.333e-17&2.730e-15$\pm$7.651e-18&1.191e-15$\pm$3.072e-17&\\
4.429e-16$\pm$1.509e-17&1.623e-16$\pm$7.323e-18&3.277e-17$\pm$7.546e-19&8.574e-18$\pm$2.922e-19&-&-\\
\\
\hline
\\
BSS4&120.00583&-10.73528&3038046598595270656&2.857e-15$\pm$9.692e-18&3.277e-15$\pm$1.009e-17&\\
-&3.339e-15$\pm$1.101e-17&-&9.181e-15$\pm$2.852e-17&1.129e-14$\pm$3.774e-17&5.751e-15$\pm$1.504e-17&\\
3.220e-15$\pm$1.579e-17&6.147e-15$\pm$5.735e-18&3.479e-15$\pm$5.793e-18&8.574e-16$\pm$2.685e-17&2.328e-15$\pm$7.469e-18&1.861e-15$\pm$4.151e-18&\\
3.229e-16$\pm$1.309e-17&1.267e-16$\pm$7.816e-18&2.728e-17$\pm$1.457e-18&7.413e-18$\pm$4.506e-19&-&-\\
\\
\hline
\\
BSS5&120.02733&-10.81018&3038043746737021440&4.175e-14$\pm$1.015e-15&4.946e-14$\pm$1.408e-15&\\
3.896e-14$\pm$7.237e-16&4.165e-14$\pm$1.487e-15&4.595e-14$\pm$3.330e-16&-&3.892e-14$\pm$1.026e-16&2.345e-14$\pm$5.981e-17&\\
-&-&1.202e-14$\pm$4.229e-17&8.603e-15$\pm$2.248e-17&6.918e-15$\pm$2.697e-17&3.059e-15$\pm$6.762e-17&\\
1.078e-15$\pm$2.582e-17&3.943e-16$\pm$1.126e-17&8.501e-17$\pm$1.801e-18&2.282e-17$\pm$5.045e-19&6.693e-19$\pm$2.700e-19&-\\
\\
\hline
\\
BSS6&120.03826&-10.80378&3038043776795821312&1.461e-13$\pm$2.300e-14&2.274e-13$\pm$7.734e-14&\\
1.979e-13$\pm$1.604e-15&1.953e-13$\pm$3.180e-14&1.340e-13$\pm$5.6e-16&-&7.654e-14$
\pm$2.018e-16&4.431e-14$\pm$1.130e-16&\\
-&-&2.134e-14$\pm$7.460e-17&-&1.169e-14$\pm$1.663e-17&4.989e-15$\pm$1.103e-16&\\
1.747e-15$\pm$4.184e-17&6.69e-16$\pm$1.725e-17&-&-&-& -
\\
\hline
\\

BSS7&119.98425&-10.69498&3038058590143955968&1.070e-14$\pm$7.041e-17&1.228e-14$\pm$1.128e-16	&\\
9.437e-15$\pm$3.583e-16&1.053e-14$\pm$7.115e-17&1.102e-14$\pm$1.656e-16&1.008e-14$\pm$4.276e-17 &8.059e-15$\pm$2.262e-17&4.895e-15$\pm$1.291e-17&\\
5.029e-15$\pm$8.755e-18&2.733e-15$\pm$4.056&2.507e-15$\pm$9.180e-18&1.789e-15$\pm$4.141e-18&1.423e-15$\pm$3.327e-18&5.867e-16$\pm$1.729e-17&\\
2.128e-16$\pm$9.210e-18&8.658e-17$\pm$6.619e-18&1.573e-17$\pm$4.201e-19&4.413e-18$\pm$1.951e-19&-&-\\
\\
\hline
\\
BSS8&120.11296&-10.76647&3038042784664224000&1.160e-15$\pm$1.270e-18&1.468e-15$\pm$2.758e-18&\\
7.909e-16$\pm$1.084e-16&1.355e-15$\pm$1.924e-18&3.641e-15$\pm$9.689e-17&7.747e-15$\pm$2.445e-17&6.288e-15$\pm$1.758e-17&4.049e-15$\pm$1.048e-17&\\
4.363e-15$\pm$7.433e-18&2.529e-15$\pm$3.498e-28&2.299e-15$\pm$8.198e-18&1.694e-15$\pm$2.030e-18&1.369e-154$\pm$5.803e-18&6.020e-16$\pm$1.830e-17&\\
2.284e-16$\pm$1.073e-17&8.674e-17$\pm$7.350e-18&1.636e-17$\pm$4.220e-19&4.226e-18$\pm$1.946e-19&-&-\\
\\
\hline
\\
BSS9&119.77495&-10.86142&3038049862770680832&1.012e-15$\pm$	7.767e-19&1.127e-15$\pm$2.263e-18&\\
7.742e-16$\pm$1.133e-16&9.929e-16$\pm$2.684e-18&2.753e-15$\pm$8.276e-17&6.78-e15$\pm$1.025e-17 &5.488e-15$\pm$1.677e-17&3.547e-15$\pm$9.140e-18&\\
3.843e-15$\pm$8.456e-18&2.243e-115$\pm$2.852e-18&2.031e-15$\pm$7.321e-18&1.504e-15$\pm$2.040e-18&1.215e-15$\pm$2.406e-18&5.326e-16$\pm$1.423e-17&\\
2.060e-16$\pm$1.120e-17&7.109e-17$\pm$7.595e-18&1.503e-17$\pm$4.016e-19&4.725e-18$\pm$1.828e-19&1.437e-18$\pm$2.314e-19&1.116e-18$\pm$1.796e-19\\
\hline
\\ 
YSS1&120.00234&-10.76055&&3.911e-16$\pm$2.094e-19&4.740e-164$\pm$5.680e-19&\\
-&3.392e-16$\pm$2.668e-19&
&1.735e-14$\pm$4.975e-17&1.500e-144$\pm$4.041e-17&1.081e-14$\pm$2.753e-17&\\1.201e-14$\pm$7.178e-18&7.942e-15$\pm$3.632e-17&
7.366e-15$\pm$2.603e-17&5.794e-15$\pm$5.091e-18&4.662e-15$\pm$1.278e-17&2.450e-15$\pm$5.190e-17&\\1.061e-15$\pm$2.835e-17&4.175e-16$\pm$1.153e-17&
8.346e-17$\pm$1.922e-18&2.274e-17$\pm$6.074e-19&-&-\\
\hline

\end{tabular}

\end{table*}

\begin{table}
\clearpage

\begin{tabular}{cccccccccccc}
\hline
\\
YSS2&120.02626&-10.77780&&2.780e-16$\pm$1.764e-19&3.500e-16$\pm$3.048e-19&\\
-&1.762e-16$\pm$6.640e-20&-&1.498e-14$\pm$3.374e-17&1.318e-14$\pm$3.581e-17&9.553e-15	2.438e-17&\\1.073e-14	1.797e-17$\pm$&7.207e-15	1.707e-17&6.623e-15$\pm$2.349e-17&5.169e-15$\pm$	1.527e-17&4.299e-15$\pm$	4.930e-18&2.280e-15$\pm$	6.091e-17&\\
9.788e-16$\pm$2.795e-17& 3.623e-16$\pm$1.368e-17&
8.194e-17$\pm$3.321e-18&
2.370e-17$\pm$9.823e-19&
1.407e-18$\pm$3.409e-19&
-\\
\hline

\\

YSS3&120.01897&-10.80028&&2.397e-16$\pm$1.128e-19&3.025e-16	$\pm$2.356e-19&\\
-&1.798e-16$\pm$9.825e-20&-&1.212e-14$\pm$2.791e-17&1.051e-14$\pm$2.863e-17&7.591e-15$\pm$1.936e-17&\\
8.416e-15$\pm$3.031e-17&5.674e-15$\pm$4.949e-18&5.205e-15$\pm$1.853e-17&4.098e-15$\pm$8.572e-18&3.322e-15$\pm$6.094e-18&
1.743e-15$\pm$4.334e-17\\
7.337e-16$\pm$1.960e-17&
2.766e-16$\pm$1.095e-17&
5.545e-17$\pm$1.175e-18&
1.575e-17$\pm$3.771e-19	&
-&-&\\

\hline
\\

RC1&119.97858&-10.7792&&2.845e-16$\pm$1.378e-19&3.414e-16	$\pm$2.410e-19&\\
-& 1.772e-16$\pm$6.958e-20&-&
-&2.494e-14$\pm$6.525e-17&	
2.201e-14$\pm$5.601e-17&\\
-&-&1.870e-14$\pm$6.577e-17&
-&-&8.987e-15$\pm$1.904e-16&\\
5.127e-15$\pm$1.228e-16&
2.098e-15$\pm$4.444e-17&
4.332e-16$\pm$9.576e-18&
1.246e-16$\pm$2.525e-18&
3.641e-18$\pm$3.186e-19&
-&\\

\hline

\\
RC2&119.924610&-10.72398&&1.094e-16$\pm$3.214e-20&1.324e-16	$\pm$5.066e-20&\\
-&8.247e-17$\pm$2.659e-20&-&-&22.348e-14$\pm$6.186e-17&2.155e-14$\pm$5.479e-17\\
-&-&1.896e-14$\pm$6.636e-17&-&-&9.341e-15$\pm$1.979e-16&\\
5.104e-15$\pm$1.222e-16&
2.213e-15$\pm$4.281e-17&
4.524e-16$\pm$9.583e-18&
1.246e-16$\pm$2.295e-18&
3.743e-18$\pm$2.827e-19&
-&\\

\hline\\

RC3&120.05109&-10.81254&&1.692e-16$\pm$5.877e-20&2.280e-16	$\pm$1.358e-19&\\
-&1.253e-16$\pm$4.726e-20&-&2.681e-14$\pm$4.340e-17&2.574e-14$\pm$6.696e-17&2.294e-14	$\pm$5.836e-17&\\
2.417e-14$\pm$0.000e+00&
1.998e-14$\pm$0.000e+00&
1.965e-14$\pm$6.899e-17&
1.658e-14$\pm$0.000e+00&
1.417e-14$\pm$7.838e-17&
9.410e-15$\pm$1.907e-16&\\

5.194e-15$\pm$1.005e-16&
2.052e-15$\pm$4.158e-17&
4.195e-16$\pm$8.886e-18&
1.145e-16$\pm$2.109e-18&
3.581e-18$\pm$2.671e-19&
-&\\

\hline
\hline

\end{tabular}
\end{table}

\begin{table*}
\caption{\label{table: Table 4}The best-fit SED parameters of BSS, YSS, and RC stars and their hot companions. For each of them, whether the single or double component SED is satisfactory in Column 2, log g in Column 3, luminosity, temperature, and radius in Columns 4--6, the reduced $\chi^{2}_{r}$ values in Column 7 (in case of double component fits, the $\chi^{2}_{r}$ values of the single fits are given in the brackets), scaling factor in Column 8, number of data points used to fit the SED is given in Column 9, and the values of vgf$_{b}$ parameter in Column 10 (in case of double component fits, the vgf$_{b}$ values of the single fits are given in the brackets).} 
	\adjustbox{max width=\textwidth}{
	\begin{tabular}{ccccccccccc}
		\hline
		\\
		Name&Component&log g&Luminosity&T$\mathrm{_{eff}}$&Radius&$\chi^{2}_{r}$&Scaling factor&N$_{fit}$&vgf$_{b}$\\
	    ~&~&~&[L$_\odot$]&[K]&[R$_\odot$]&($\chi^{2}_{r,single}$)&~&~&(vgf$_{b,single}$)
		 \\
		\hline 
		\hline
		\\
BSS1&A&3.5&31.18$\pm$9.05&7750$\pm$125&3.10$\pm$0.44&308.06 (649)&5.06E-22&16&0.15 (0.25) \\	

~&B&7.5&0.44$^{+0.22}_{-0.19}$&13250$\pm$250&0.13$\pm+0.02$&-&8.24E-25&-&	\\
\hline
\\

BSS2&A&3.0&19.30$\pm$5.60&8000$\pm$125&2.29$\pm$0.33&266.26 (694)&2.77E-22&14&0.03 (0.16) \\                    

~&B&8.0&0.45$^{+0.23}_{-0.20}$&15000$\pm$250&0.10$\pm$0.01&-&5.21E-25&-& \\
\hline
\\

BSS3&single&4.0&31.58$\pm$9.16&9250$\pm$125&2.19$\pm$0.31&48.32&2.53E-22&18&0.58  \\ 
\hline

\\

BSS5&single&4.5&84.37$\pm$24.49&9500$\pm$125&3.38$\pm$0.49&13.26&6.03E-22&14&0.54 \\  
\hline
\\

BSS7&single&4.5&18.40$\pm$5.34&9750$\pm$125&1.51$\pm$0.21&68.83&1.21E-22&17&0.57 \\
\hline
\\

BSS8&A&3.0&12.06$\pm$3.50&8250$\pm$125&1.70$\pm$0.24&596.22(1864)&1.52E-22&18& 0.91 (1.51)\\ 

~&B&7.5&0.24$^{+0.12}_{-0.11}$&19000$\pm$250&0.05$\pm$0.01&-&1.19E-25&-& \\
\hline
\\

BSS9&single&3.0&10.58$\pm$3.07&8250$\pm$125&1.60$\pm$0.23&2389&1.35E-22&16&1.59 \\
\hline
\\

YSS1&A&3.5&30.73$\pm$8.92&6750$\pm$125&4.06$\pm$0.58&17.88 (4771)&8.69E-22&16&0.04(2.09) \\
~&B&4.5&0.14$^{+0.08}_{-0.07}$&29000$\pm$1000&0.01$\pm$0.00&-&1.15E-26& &  &\\
\hline
\\

YSS2&single&3.0&26.72$\pm$7.77&6500$\pm$125&4.08$\pm$0.59&29.39&8.77E-22&17&0.66 \\
\hline
\\

YSS3&A&3.5&21.84$\pm$6.34&6750$\pm$125&3.42$\pm$0.49&15.31 (33835)&6.16E-22&16&0.10 (2.13)\\

~&B&3.5&0.07$^{+0.04}_{-0.03}$&26000$\pm$1000&0.01$\pm$0.00&-&8.48E-27&-& \\
\hline
\\

RC1&A&4.0&67.49$\pm$19.67&5250$\pm$125&9.86$\pm$1.42&28.12 (1140)&5.11E-21&13&3.07 (91) \\
~&B&5.0&0.13$^{+0.08}_{-0.06}$&30000$\pm$1000&0.01$\pm$0.00&-&9.39E-27&-& \\
\hline
\\

RC2&A& 3.5&67.54$\pm$19.69&5000$\pm$125&10.96$\pm$1.58&71.5 (517632) & 6.31E-21&13&0.25 (0.99)  \\
~& B &5.0&0.05$^{+0.03}_{-0.02}$&25000$\pm$1000&0.01$\pm$0.00&-&8.48E-27&-& \\
\hline
\\

RC3&A&5.0&69.30$\pm$20.42&5250$\pm$125&10.04$\pm$1.45&  17.88 (26880) &5.29E-21&18&0.04(1.33) \\
~&B&5.0&0.09$^{+0.05}_{-0.04}$&31000$\pm$1000&0.01$\pm$0.00&-& 5.64E-27& & & \\
\hline
\hline
		\label{table4}
\end{tabular}
}

\end{table*}

\bsp	
\label{lastpage}
\end{document}